\documentclass[preprint,aps,unsortedaddress]{revtex4}
\usepackage{graphicx}
\usepackage{epstopdf}
\usepackage{dcolumn}
\usepackage{amsmath}
 \usepackage{multirow}
\usepackage{adjustbox}
\usepackage[normalem]{ulem}
\usepackage[usenames, dvipsnames]{color}
\newcommand{\be}{\begin{equation}}
\newcommand{\ee}{\end{equation}}
\newcommand{\bea}{\begin{eqnarray}}
\newcommand{\eea}{\end{eqnarray}}

\newcommand{\bfk}{\mbox{\boldmath $k$}}
\newcommand{\bfq}{\mbox{\boldmath $q$}}

\newcommand{\pup}{p^\uparrow}

\newcommand{\bfS}{\mbox{\boldmath $S$}}

\def\lsim{\mathrel{\rlap{\lower4pt\hbox{\hskip1pt$\sim$}}\raise1pt\hbox{$<$}}}
\def\gsim{\mathrel{\rlap{\lower4pt\hbox{\hskip1pt$\sim$}}\raise1pt\hbox{$>$}}}

\begin{document}

\title{Transverse Single Spin Asymmetry in $p+	p^\uparrow \rightarrow J/\psi +X$}
\author{Rohini M. Godbole}
\affiliation{Centre for High Energy Physics, Indian Institute of Science, Bangalore, India.}
\email{rohini@chep.iisc.ernet.in}
\author{Abhiram Kaushik}
\affiliation{Centre for High Energy Physics, Indian Institute of Science, Bangalore, India.}
\email{abhiramb@chep.iisc.ernet.in}
\author{Anuradha Misra}
\affiliation{Department of Physics, University of Mumbai, Mumbai, India}
\email{misra@physics.mu.ac.in}
\author{Vaibhav Rawoot}
\affiliation{Department of Physics, University of Mumbai, Mumbai, India.}
\email{vaibhavrawoot@gmail.com}
\author{Bipin Sonawane}
\affiliation{Department of Physics, University of Mumbai, Mumbai, India.}
\email{bipin.sonawane@physics.mu.ac.in}

\date{\today}

\begin{abstract}
We present estimates of transverse single spin asymmetry (TSSA) in $p+p^\uparrow \rightarrow J/\psi+X$ within  the colour evaporation model of charmonium production in a generalized parton model (GPM) framework, using the recently obtained best fit parameters for the gluon Sivers function (GSF) extracted from PHENIX data on TSSA in $p+p^\uparrow \to \pi^0+X$ at midrapidity. We calculate asymmetry at $\sqrt{s} = 200$ GeV, and compare the results with PHENIX data on TSSA in the process $p + p^\uparrow \to J/\psi+X$.  We also present estimates for asymmetry at $\sqrt{s} = 115 $ GeV corresponding to the  proposed fixed target experiment AFTER@LHC and at   $\sqrt{s} = 500$ GeV corresponding to the higher RHIC energy. Finally, we investigate the effect of the transverse momentum dependent (TMD) evolution of the densities involved, on the asymmetry.
\end{abstract}

\preprint{\bf CERN-TH-2017-047}
\maketitle

\section{Introduction}

The role of intrinsic transverse momentum and spin distribution of partons inside the hadrons in explaining the azimuthal asymmetries arising in experiments 
involving polarized beams ~\cite{Adare:2013ekj, Klem:1976ui,Antille:1980th} has been a subject of keen interest during  the past decade. Transverse Single Spin Asymmetries (TSSA's),  measured in  meson production in hadronic collisions as well as  in semi inclusive deep inelastic scattering (SIDIS) experiments, provide useful information towards understanding the transverse dynamics of partons inside the hadrons.  

It is  known for long~\cite{Kane:1978nd} that the conventional QCD factorization with collinear parton distribution functions (PDFs) and fragmentation functions (FFs) is not sufficient to account for the single spin asymmetries (SSAs) observed experimentally. It is now well established that the  
SSAs observed  in the hadroproduction of mesons and semi inclusive deep inelastic 
scattering (SIDIS)~\cite{Airapetian:2009ae,Alekseev:2008aa,Adolph:2012sp,Qian:2011py} 
can be explained in terms of orbital motion of quarks and gluons and the spin structure of nucleons. One of the two approaches towards building up a suitable framework 
involves generalization of the concept of collinear PDFs by including the transverse momentum and spin dependence of the partons into  PDFs and FFs,  which are then collectively 
referred to as TMDPDFs. This approach known as TMD approach has been successful in explaining the existing data in some processes, for example in inclusive pion production in $pp$ collisions~\cite{D'Alesio:2004up} and $Z$-boson production in $pp$ and $p{\bar p} $ collisions~\cite{Echevarria:2014xaa}. 
In the Transverse Momentum Dependent factorization scheme~\cite{Collins:2011zzd, Ji:2004xq, Ji:2004wu, Bacchetta:2008xw}, 
the hadronic cross-section is then expressed  as a convolution of TMDPDFs, TMD FFs and partonic cross-section.

One of the interesting TMDPDFs, which has been a subject of a large number of studies is the Sivers Function, which can be interpreted as the number density of unpolarized quarks or gluons inside a transversely polarized proton. Information on the Sivers function can shed light on the 3-dimensional structure of nucleon and may also provide an estimate of the orbital angular momentum of quarks and gluons~\cite{Boer:2011fh, Bacchetta:2011gx}. It was introduced by Sivers in Ref.~\cite{Sivers:1989cc,Sivers:1990fh} wherein a  transverse momentum dependent PDF, now known as  Sivers function,  was introduced for the first time to account for the large asymmetries observed in $p\pup$ scattering~\cite{Sivers:1989cc}. Quark Sivers function, which is  usually coupled to reasonably well parametrized unpolarized fragmentation functions, has been one of the first  TMDs extracted from data~\cite{Boer:2011fh}. 
 All the known information on quark Sivers function has been obtained from SIDIS data and $\gamma^*-N $ scattering in the   centre of mass frame.
   Initial  extractions of up and down quark Sivers functions,  from  HERMES and COMPASS data,  were obtained using parameterizations that did not take into account sea quarks~\cite{Anselmino:2010bs,Anselmino:2005ea}. Later, magnitude of asymmetry for $K^+$ measured at HERMES indicated the possible contribution of sea quarks and fits were obtained taking into account contributions of sea quarks also~\cite{Anselmino:2010bs}. Parametrizations of Sivers function for up and down quarks have been found to be in agreement with light-cone models~\cite{Bacchetta:2008af} and a quark-diquark spectator model~\cite{Gamberg:2007wm}.  Extractions of quark Sivers function taking into account TMD evolution have  also been obtained by performing a global fitting of all data on Sivers asymmetry in SIDIS from HERMES, COMPASS and Jefferson Lab~\cite{Echevarria:2014xaa} and  based on these, predictions have been made for Sivers asymmetry in the DY process and $W$-boson production. 
Although there are many parameterizations available for quark Sivers function, not much information is available on gluon Sivers function (GSF). Some of  processes which  have been proposed for obtaining information about GSF are $p\pup \rightarrow \gamma+X$~\cite{Schmidt:2005gv}, $p\pup \rightarrow D+X$~\cite{Anselmino:2004nk,  Kang:2008qh}$, p\pup \rightarrow \gamma  +\text{jet}+X $\cite{Schmidt:2005gv,Bacchetta:2007sz}, $p\pup \rightarrow \gamma^*+X \rightarrow \mu^+ \mu^-+ X $~\cite{Schmidt:2005gv} and $p\pup \rightarrow \eta_{c/b}+X$~\cite{Schafer:2013wca}.

Heavy flavour production- both open and closed- are considered to be clean probes of the GSF since heavy quarks are predominantly produced via gluon-gluon fusion and thus can be used to isolate gluon dynamics within hadrons~\cite{Yuan:2008vn, Brodsky:2002cx,Brodsky:2002rv, Anselmino:2004nk,Godbole:2016ixc,DAlesio:2017rzj}. The possibility of getting information  on the GSF by looking at  $D$-meson production in polarized proton-proton scattering at RHIC  has been discussed by Anselmino {\it et al.}~\cite{Anselmino:2004nk} using a saturated GSF. The first phenomenological study on the gluon Sivers function has recently been performed by  D'Alesio, Murgia and Pisano~\cite{D'Alesio:2015uta}, wherein the gluon Sivers function has been fitted to midrapidity data on transverse SSA in $pp\rightarrow \pi^0+X$ measured by PHENIX collaboration at RHIC~\cite{Adare:2013ekj}. In our previous work,  where we made predictions for TSSA in electroproduction of $J/\psi $ assuming a transverse momentum dependent factorization within colour evaporation model (CEM) of charmonium production, we had used parameterization of the gluon Sivers function suggested by Boer and Vogelsang~\cite{Boer:2003tx},  in which the $x$-dependence of the GSF was modeled on that of the $u$ and $d$ quark Sivers functions. We will call these BV parameters in the following.  In our recent work on TSSA  in $D$-meson production~\cite{Godbole:2016tvq}, we have  used the directly fitted GSF parameters of Ref.~\cite{D'Alesio:2015uta} and have compared the estimates using these with the results obtained using BV parameters.

In this work, we present predictions for TSSA in the process $p\pup \rightarrow J/\psi+X$ using recent directly fitted parameters, which we will call DMP fits \cite{D'Alesio:2015uta}. We compare these results with predictions obtained using the BV parameters, which are based on experimentally fitted  quark Sivers parameters \cite{Boer:2003tx}. We then compare our results with the recent measurements of Sivers asymmetry at PHENIX experiment in $J/\psi$ production~\cite{Adare:2010bd}.  A similar comparison with PHENIX results  has been performed recently in Ref.~\cite{DAlesio:2017rzj} using colour singlet model (CSM) and a maximized GSF, whereas in our work we use CEM and compare various parameterizations of GSF available. We also present estimates of asymmetry for future proposed experiments at AFTER@LHC which is a fixed target experiment with  $\sqrt{s} = 115$ GeV and for  $\sqrt{s} = 500$ GeV  which will be explored at RHIC. Finally, to assess the effect of QCD evolution on asymmetry, we  compare the predictions based on DGLAP and TMD evolution of the unpolarized TMDPDF and gluon Sivers function. This comparison is performed  using the BV parameters because direct fits of the GSF with TMD evolution taken into account are currently not available.
\section{Formalism}
\subsection{TSSA in the process  $p+\pup \to J/\psi+X$ using  colour evaporation model}
We consider the transverse single spin asymmetry,
\begin{equation} 
A_N = \frac{d\sigma^\uparrow  -  d\sigma^\downarrow}{d\sigma^\uparrow  + d\sigma^\downarrow}
\label{an} 
\end{equation} 
for  the process $p+\pup \to J/\psi+X$ using the colour evaporation model ~\cite{GayDucati:1999kh} of $J/\psi$ production.

In the colour evaporation model, the total cross-section for the production of $J/\psi$ at leading order (LO) is proportional to the rate of $c\bar{c}$ production integrated over the invariant mass-squared range $4m_c^2$ to $4m_D^2$, where $m_c$ is the mass of the charm quark and $m_D$ is the open charm threshold~\cite{Halzen:1977im}:
\bea
\sigma^{p+p \to J/\psi + X} = F_{J/\psi} \int_{4m^2_c}^{4m^2_D} dM^2_{c\bar c}
\int d x_a d x_b \bigg[f_{g/p}(x_a) f_{g/p}(x_b) \frac{d\hat\sigma^{gg \to c\bar c}}{dM^2_{c\bar c}} \nonumber \\
+ \sum_q f_{q/p}(x_a) f_{{\bar q}/p}(x_b) \frac{d\hat\sigma^{q{\bar q} \to c\bar c}}{dM^2_{c\bar c}}\bigg],
\label{cem-xs}
\eea
where $q = u, d, s, \bar u, \bar d, \bar s$. 

Here, the CEM parameter $F_{J/\psi}$ is the fraction that gives the probability of $J/\psi$ production below $D\bar D$ threshold. 

Here, we use a phenomenological approach referred to in literature as the Generalized Parton Model (GPM), which has been used to estimate SSAs in several  processes like $p\pup \rightarrow D+X$~\cite{Anselmino:2004nk, Godbole:2016tvq,DAlesio:2017rzj}, $ p\pup \rightarrow \gamma  +\text{jet}+X $\cite{Schmidt:2005gv,Bacchetta:2007sz} and $p\pup \rightarrow \pi +X$~\cite{D'Alesio:2004up} for which TMD factorization has not yet been established. A rigorous treatment  will require inclusion of intrinsic transverse momentum effects through a consideration of higher twist effects. However, motivated by the phenomenological successes of the GPM~\cite{DAlesio:2007bjf,D'Alesio:2004up}, we assume  a generalization of  CEM expression and  include TMDPDFs  thus expressing the cross section for the transversely polarized scattering process $p + p^\uparrow \to J/\psi + X$ as 
\bea
\sigma^{p+p^\uparrow \to J/\psi + X} = F_{J/\psi} \int_{4m^2_c}^{4m^2_D} dM^2_{c\bar c}
\int d x_a d^2 \bfk_{\perp_a} d x_b d^2 \bfk_{\perp_b} 
\Biggl\{f_{g/p}(x_a, \bfk_{\perp_a}) f_{g/p^\uparrow}(x_b, \bfk_{\perp_b}) \frac{d\hat\sigma^{gg\to c\bar c}}{dM^2_{c\bar c}} \nonumber \\
+ \sum_q \biggl[f_{q/p}(x_a, \bfk_{\perp_a}) f_{\bar{q}/p^\uparrow}(x_b, \bfk_{\perp_b}) \frac{d\hat\sigma^{q\bar{q}\to c\bar c}}{dM^2_{c\bar c}}\biggr]\Biggr\},
\label{tmd-cem}
\eea
where $q = u, d, s, \bar u, \bar d, \bar s$ and the gluon and quark densities have been replaced by transverse momentum dependent gluon and quark PDFs. In Eq.~\ref{tmd-cem}, $f_{g/p^{\uparrow(\downarrow)}}(x,\bfk_\perp)$ is the TMDPDF describing the distribution of gluons in proton which is  transversely polarized w.r.t the beam axis with the polarization being upwards (downwards) with respect to the production plane. For a general value of the transverse spin $\bfS_\perp$, it is parametrised in terms of the gluon Sivers function (GSF) $\Delta^{N}f_{g/p^\uparrow}$, as follows:
\bea
f_{g/p^\uparrow}(x,\bfk_\perp,\bfS_\perp;Q)&=&f_{g/p}(x,k_\perp;Q)-f^{\perp i}_{1T}(x,k_\perp;Q)\frac{\epsilon_{ab}k_\perp^aS_\perp^b}{M_p}
\\\nonumber&=&f_{g/p}(x,k_\perp;Q)+\frac{1}{2}\Delta^N f_{g/p^\uparrow}(x,\bfk_\perp,\bfS_\perp;Q)
\\\nonumber&=&f_{g/p}(x,k_\perp;Q)+\frac{1}{2}\Delta^N f_{g/p^\uparrow}(x,k_\perp;Q)\frac{\epsilon_{ab}k_\perp^a S_\perp^b}{k_\perp}.
\label{Sivers}
\eea
Any non-zero TSSA in the process considered would primarily arise due to an azimuthal anisotropy in the distribution of gluon transverse momenta in the polarized proton. This anisotropy is parametrised by the gluon Sivers distribution.

Following Ref.~\cite{Godbole:2012bx} we can then write the numerator and denominator of Eq. \ref{an} as, 

\begin{align}
&&\frac{d^{3}\sigma^\uparrow}{dyd^2\bfq_T}-\frac{d^3\sigma^\downarrow}{dyd^2\bfq_T} = 
\frac{F_{J/\psi}}{s}\int [dM_{c\bar{c}}^2d^2\bfk_{\perp a}d^2\bfk_{\perp b}]
\delta^2(\bfk_{\perp a}+\bfk_{\perp b}-\bfq_T) \nonumber \\
&& \times\>  \Biggl\{\Delta^{N}f_{g/p^\uparrow}(x_{a},\bfk_{\perp a}) f_{g/p}(x_{b},\bfk_{\perp b}) 
\hat\sigma_{0}^{g g\rightarrow c\bar{c}}(M_{c\bar{c}}^2) \nonumber \\
&& +\sum_q \biggl[\Delta^{N}f_{q/p^\uparrow}(x_{a},\bfk_{\perp a}) f_{\bar q/p}(x_{b},\bfk_{\perp b}) 
\hat\sigma_{0}^{q\bar q\rightarrow c\bar{c}}(M_{c\bar{c}}^2) \biggr] \Biggr\}
\label{num-ssa}
\end{align}
and
\begin{align}
&&\frac{d^{3}\sigma^\uparrow}{dyd^2\bfq_T}+\frac{d^3\sigma^\downarrow}{dyd^2\bfq_T}=  
\frac{2 F_{J/\psi}}{s}\int [dM_{c\bar{c}}^2d^2\bfk_{\perp a}d^2\bfk_{\perp b}]
\delta^2(\bfk_{\perp a}+\bfk_{\perp b}-\bfq_T) \nonumber \\
&& \times\> \Biggl\{f_{g/p^\uparrow}(x_{a},\bfk_{\perp a}) f_{g/p}(x_{b},\bfk_{\perp b}) 
\hat\sigma_{0}^{g g\rightarrow c\bar{c}}(M_{c\bar{c}}^2) \nonumber \\
&& +\sum_q \biggl[f_{q/p^\uparrow}(x_{a},\bfk_{\perp a}) f_{\bar q/p}(x_{b},\bfk_{\perp b}) 
\hat\sigma_{0}^{q\bar q\rightarrow c\bar{c}}(M_{c\bar{c}}^2) \biggr] \Biggr\}
\label{deno-ssa}
\end{align}

with,
\bea
x_{a,b}=\frac{M_{c\bar{c}}}{\sqrt{s}}e^{\pm y}.
\label{xy}
\eea
Here, $y$ and $\mathbf{q}_T$ are the rapidity and transverse momentum of the $J/\psi$ and we consider the plane of production of the $J/\psi$ to be perpendicular to the proton spin $\bfS_\perp$. The partonic cross-sections for production of a $c\bar{c}$ pair of mass $M$ are  given by~\cite{Gluck:1978bf}
\be
\hat{\sigma_0}^{g g\rightarrow c\bar{c}}(M^2)=
\frac{\pi \alpha_s^2}{3\hat{s}}
\left[\left(1+ v +\frac{1}{16} v^2\right)\ln{\frac{1+\sqrt{1- v}}{1-\sqrt{1-v}}}
-\left(\frac{7}{4}+\frac{31}{16}v\right)\sqrt{1-v}\right]
\label{gg2ccbarxs}
\ee
and 
\be
\hat{\sigma_0}^{q \bar q \rightarrow c\bar{c}}(M^2)=
\frac{2}{9}\left(\frac{4 \pi \alpha_s^2}{3\hat{s}}\right)
\left(1 + \frac{1}{2} v\right)\sqrt{1-v},
\label{qqbar2ccbarxs}
\ee
where $v=\frac{4 m_c^2}{M^2}$. 


\subsection{Parametrization of the TMDs}
For the predictions with the two DMP fits~\cite{D'Alesio:2015uta}, we adopt the same functional forms for the TMDs using which they were extracted. We use the standard form for the unpolarized TMDPDF with a factorized Gaussian $k_\perp$-dependence, 
\be
f_{i/p}(x,k_\perp;Q)=f_{i/p}(x,Q)\frac{1}{\pi\langle k_\perp^2\rangle}e^{-k_\perp^2/\langle k_\perp^2\rangle}
\label{unptmd}
\ee
with $\langle k_{ \perp }^2 \rangle$ = 0.25 GeV$^2$.
The Sivers function is parameterized as~\cite{Anselmino:2008sga}
\be
\Delta^N f_{i/p^\uparrow}(x,k_\perp;Q)=2\mathcal{N}_{i}(x)f_{i/p}(x,Q)h(k_\perp)\frac{e^{-k^2_\perp/\langle k_\perp^2\rangle}}{\pi \langle k_\perp^2\rangle}
\label{param-siv}
\ee
with, 
\be
\mathcal{N}_i(x)=N_i x^{\alpha_i}(1-x)^{\beta_i}\frac{(\alpha_i+\beta_i)^{\alpha_i+\beta_i}}{\alpha_i^{\alpha_i} \beta_i^{\beta_i}}
\label{Nx}
\ee
and
\be
h(k_\perp)=\sqrt{2e}\frac{k_\perp}{M_1}e^{-k_\perp^2/M_1^2},
\ee
where $\mathcal{N}_i$, $\alpha_i$, $\beta_i$ and $M_1$ are all parameters determined by fits to data and $e$ is Euler's number.

As mentioned in the introduction, the two DMP extractions of the GSF, namely SIDIS1 and SIDIS2 were obtained by fitting to data on TSSA in $p^\uparrow p\to\pi^0+X$ at RHIC with quark Sivers function (QSFs) extracted earlier from SIDIS data being used to account for the quark contribution to $A_N$. 
In obtaining the SIDIS1 fits, only the $u$ and $d$ flavour were taken into account, using the data on pion production from the HERMES experiment and positive hadron production from the COMPASS experiment. SIDIS2 parameters were obtained using flavour segregated data on pion and kaon production so here all three light flavours were taken into account.  Furthermore, the QSFs used in the SIDIS1 fit were obtained with the set of fragmentation functions by Kretzer~\cite{Kretzer:2000yf} and those of SIDIS2 were obtained with the set by de Florian, Sassot and Stratmann (DSS)~\cite{deFlorian:2007aj}. The values of the parameters of the two fits are given in Table~\ref{SIDIS-gluon-fits}. We give predictions for TSSA using these.

\begin{table}[ht]
\centering
\begin{tabular}{|l|l|l|l|l|l|l|}
\hline
SIDIS1 & \multicolumn{2}{l|}{$N_g=0.65$} & $\alpha_g=2.8$ & $\beta_g=2.8$ & $\rho=0.687$ & \multirow{2}{*}{$\langle k^2_\perp\rangle=0.25 GeV^2$} \\ \cline{1-6}
SIDIS2 & \multicolumn{2}{l|}{$N_g=0.05$} & $\alpha_g=0.8$ & $\beta_g=1.4$ & $\rho=0.576$ &                                                        \\ \cline{1-7}
\end{tabular}
\caption{DMP fit parameters. $\rho=M_1^2/(\langle k_\perp^2\rangle+M_1^2)$.}
\label{SIDIS-gluon-fits}
\end{table}

\subsection{TMD Evolution}
The QCD evolution formalism of the unpolarised TMD and the Sivers function has been obtained for DY and SIDIS~\cite{Echevarria:2014xaa} both of which involve a color singlet photon. One expects the TMD evolution of TMDs to be different for more complicated processes. However, since $J/\psi$ is a color singlet, in this case one can assume the TMD evolution of Ref.~\cite{Echevarria:2014xaa} for a preliminary assessment of the  effect of evolution on asymmetries. A more rigorous approach to TMD evolution for quarkonium will be closely related to the issue of validity of TMD factorization for quarkonium which is not yet established.

The energy evolution of a TMDPDF $F(x,k_\perp;Q)$ is best described through its Fourier transform into coordinate space which is given by 
\be
F(x,b;Q)=\int d^2k_\perp e^{-i\vec{k}_\perp.\vec{b}_\perp}F(x,k_\perp;Q).
\ee
The evolution of $b$-space TMDPDFs can then be written as,
\be
F(x,b;Q_f)=F(x,b,Q_i)R_{pert}(Q_f,Q_i,b_*)R_{NP}(Q_f,b),
\label{eee}
\ee
where, $R_{pert}$ is the perturbatively calculable part of the evolution kernel, $R_{NP}$ is a nonperturbative Sudakov factor and $b_*=b/\sqrt{1+(b/b_\text{max})^2}$ is used to stitch together the perturbative part of the kernel,  which is valid for $b<<b_{max}$,  with the nonperturbative part, which is valid for large $b$. Following Ref.~\cite{Echevarria:2014xaa}, we choose an initial scale $Q_i=c/b_*$ for TMD evolution. Here $c=2e^{-\gamma_E}$ where $\gamma_E$ is the Euler-Mascheroni constant.

Setting $Q_i=c/b_*$ and $Q_f=Q$, the perturbative part can be written as,
\bea
R_{pert}= \exp\left\{-{\int_{c/b^*} ^{Q} \frac {d\mu}{\mu}\left(A\ln\frac{Q^2}{\mu^2}+B\right)}\right\},
\eea

where $A=\Gamma_\text{cusp}$ and $B=\gamma^V$ are anomalous dimensions that can be expanded perturbatively. The expansion coefficients with the appropriate gluon anomalous dimensions at NLL are~\cite{Echevarria:2014xaa}
\bea
A^{(1)}&=&C_A,
\\
A^{(2)}&=&\frac{1}{2}C_A\left(C_A\left(\frac{67}{18}-\frac{\pi^2}{6}\right)-\frac{5}{9}C_AN_f\right),
\\
B^{(1)}&=&-\frac{1}{2}\left(\frac{11}{3}C_A-\frac{2}{3}N_f\right).
\eea
The nonperturbative part of the evolution kernel, $R_{NP}$ is
\bea
R_{NP}=\exp\left\{-b^{2}\left(g_1^\text{TMD}+\frac{ g_2}{2}\ln\frac{Q}{Q_0}\right)\right\},
\eea
where, $g_2$ is a factor which takes the same value for all quark TMDPDFs~\cite{Echevarria:2014xaa,Collins:2011zzd}
 and $g_1$ is TMDPDF specific and is proportional to the intrinsic transverse momentum width of the particular TMDPDF at the momentum scale $Q_0$. In case of gluon TMDPDFs, $g_2$ is to be multiplied by a factor of $\frac{C_A}{C_F}$~\cite{Kang:2017glf}. Assuming a factorized Gaussian form at scale $Q_0$, we have
\bea
g_1^{\rm unpol} = \frac{\langle k_\perp^2\rangle_{Q_0}}{4}
\qquad\text{and}\qquad
g_1^{\rm Sivers} = \frac{\langle k_{s\perp}^2\rangle_{Q_0}}{4}.
\eea

Expanding the TMDPDF $F(x,b;Q)$ at the initial scale in terms of its corresponding collinear density at leading order (LO), for the unpolarized TMDPDF we get,
\bea
f_{i/p}(x,b;Q)&=&f_{i/p}(x,c/b_*)\exp\left\{-{\int_{c/b^*} ^{Q} \frac {d\mu}{\mu}\left(A\ln\frac{Q^2}{\mu^2}+B\right)}\right\}
\\
\nonumber&\times&\exp\left\{-b^{2}\left(g_1^\text{unpol}+\frac{g_2}{2}\ln\frac{Q}{Q_0}\right)\right\}.
\label{fbT}
\eea

In the case of the Sivers function, the evolution of its derivative in $b$-space can be written in the form of Eq.~\ref{eee} leading to,
\bea
f'^{\perp i}_{1T}(x,b;Q)&=&\frac{M_p b}{2}T_{i,F}(x,x,c/b_*)\exp\left\{-{\int_{c/b^*} ^{Q} \frac {d\mu}{\mu}\left(A\ln\frac{Q^2}{\mu^2}+B\right)}\right\}
\\
\nonumber&\times&\exp\left\{-b^{2}\left(g_1^\text{Sivers}+\frac{g_2}{2}\ln\frac{Q}{Q_0}\right)\right\},
\label{fSbT}
\eea
where, $f'^{\perp}_{1T}(x,b;\mu)\equiv\frac{\partial f^\perp_{1T}}{\partial b}$. Here the Qiu-Sterman function for parton $i$, $T_{i,F}(x,x,\mu)$ is obtained when expanding $f'^{\perp}_{1T}(x,x,\mu)$ at LO and can be parametrized as
\be
T_{i,F}(x,x,\mu)=\mathcal{N}_i(x)f_{i/p}(x,\mu)
\label{QS}
\ee
with $\mathcal{N}_i(x)$ having the same form as in Eq.~\ref{Nx}.

The expressions for the TMDs in $k_\perp$-space can be obtained by Fourier transforming the $b$-space expressions:
\bea
f_{i/p}(x,k_\perp;Q)&=&\frac{1}{2\pi}\int^\infty_0db~bJ_0(k_\perp b)f_{i/p}(x,b;Q)
\\
f^{\perp i}_{1T}(x,k_\perp;Q)&=&\frac{-1}{2\pi k_\perp}\int^\infty_0db~bJ_1(k_\perp b)f'^{\perp i}_{1T}(x,b;Q).
\eea
The above expression for the Sivers function is related to $\Delta^Nf_{i/p^\uparrow}(x,k_\perp;Q)$ through Eq.~4.

For the purpose of studying the effect of the transverse-momentum-dependent evolution of the densities on the asymmetry predictions, we need to use  the BV models for the GSF since there are no available fits of the GSF that take into account TMD evolution. We therefore consider the following two models of the GSF wherein the $x$-dependent term $\mathcal{N}_g(x)$, is modeled on that of the quarks:
\begin{enumerate}
\item BV (A): $\mathcal{N}_g(x)=\{\mathcal{N}_u(x)+\mathcal{N}_d(x)\}/2$
\item BV (B): $\mathcal{N}_g(x)=\mathcal{N}_d(x)$
\end{enumerate}
where both $\mathcal{N}_u(x)$ and $\mathcal{N}_d(x)$ are of the form given in Eq.~\ref{Nx}.
These models were first used in Ref.~\cite{Boer:2003tx} by Boer and Vogelsang and hence we will refer to these as the BV models.

For the predictions with TMD evolved densities using the above two models, we use the following set of parameters given in Ref.~\cite{Echevarria:2014xaa},
\begin{table}[ht]
\centering
\label{my-label}
\begin{tabular}{|l|l|l|l|}
\hline
$N_u=0.106$               & $\alpha_u=1.051$             & $\beta_u=4.857$ & $\langle k_\perp^2\rangle_{Q_0}=0.38$ GeV$^2$     \\ \hline
$N_d=-0.163$              & $\alpha_d=1.552$             & $\beta_d=4.857$ & $\langle k_{s\perp}^2\rangle_{Q_0}=0.282$ GeV$^2$ \\ \hline
\multicolumn{2}{|l|}{$b_\text{max}=1.5$ GeV$^{-1}$} & \multicolumn{2}{l|}{$g_2=0.16$ GeV$^2$}                        \\ \hline
\end{tabular}
\caption{Parameters for TMD evolved densities from Ref.~\cite{Echevarria:2014xaa}.}
\end{table}

The predictions made with the TMD evolved densities using the BV models are compared with those obtained with DGLAP evolved densities using the same models with the following set of parameters given in ~\cite{Anselmino:2011gs}
\begin{table}[ht]
\centering
\begin{tabular}{|l|l|l|l|}
\hline
$N_u=0.4$   & $\alpha_u=0.35$ & $\beta_u=2.6$ & $\langle k_\perp^2\rangle=0.25$ GeV$^2$ \\ \hline
$N_d=-0.97$ & $\alpha_d=0.44$ & $\beta_d=0.9$  & $M_1^2=0.19$                                  \\ \hline
\end{tabular}
\caption{Parameters for DGLAP evolved densities from ~\cite{Anselmino:2011gs}.}
\label{my-label2}
\end{table}

\section{Results}

In this section, we present predictions of transverse single spin asymmetry in $p + p^\uparrow \rightarrow J/\psi+X$, obtained using the DMP fits~\cite{D'Alesio:2015uta} and the BV models~\cite{Boer:2003tx} of the GSF and corresponding best fit parameters of QSFs. The QSFs corresponding to SIDIS-1 and SIDIS-2 are given in ref.~\cite{Anselmino:2005ea} 
and ref.~\cite{Anselmino:2008sga} respectively. The best fits of QSF corresponding to BV models 
are given in ref.~\cite{Anselmino:2011gs} for DGLAP evolved  densities and in ref.~\cite{Echevarria:2014xaa} for the  TMD evolved ones. Our predictions of TSSA are given for three different centre of mass energies $\sqrt{s}=115$ GeV (AFTER@LHC), 200 GeV (RHIC1) and 500 GeV (RHIC2).
We present asymmetry predictions as a function of: (i) the transverse momentum $q_T$, with the rapidity integrated over $-2.8\leq y \leq 0.2$ for $ \sqrt{s}=115$ GeV, $2\leq y\leq 3 $ and $ 3\leq y \leq 3.8 $ for $\sqrt{s}=200$ GeV,  and $2\leq y \leq 3 $ and $3 \leq y \leq 4 $ for $\sqrt{s}=500$ GeV (ii) the rapidity $y$, with the transverse momentum integrated in the range $0<q_T<1.4$ GeV. The given rapidity ranges were chosen keeping in mind the proposed fsPHENIX ~\cite{Barish:2012ha, Aschenauer:2015eha} upgrade which will bring the forward coverage of the detector to $1\leq\eta\leq4$. For convenience, we will refer to (i) and (ii) as $q_T$-asymmetry and $y$-asymmetry respectively.  We then present a comparison of asymmetries estimated using DGLAP evolved  densities with those obtained using TMD-evolved densities in order to study the effect of TMD evolution. 
We then compare the asymmetries  obtained using the aforementioned fits and models, with measurements of TSSA in $p + p^\uparrow \rightarrow J/\psi+X$ at $\sqrt{s}=200$ GeV, performed by the PHENIX collaboration at RHIC~\cite{Adare:2010bd}. These measurements were performed in the forward ($1.2\leq y\leq 2.2$), backward ($-2.2\leq y\leq-1.2$) and midrapidity ($-0.35\leq y\leq0.35$) regions with $0\leq q_T\leq1.4$ GeV. Finally, we consider the possibility of probing the asymmetry in the extended forward region ($2.2\leq y\leq4.0$) that will become accessible under the proposed fsPHENIX upgrade~\cite{Barish:2012ha, Aschenauer:2015eha}.

\begin{figure}[ht]
\begin{center}
\includegraphics[width=0.49\linewidth,angle=0]{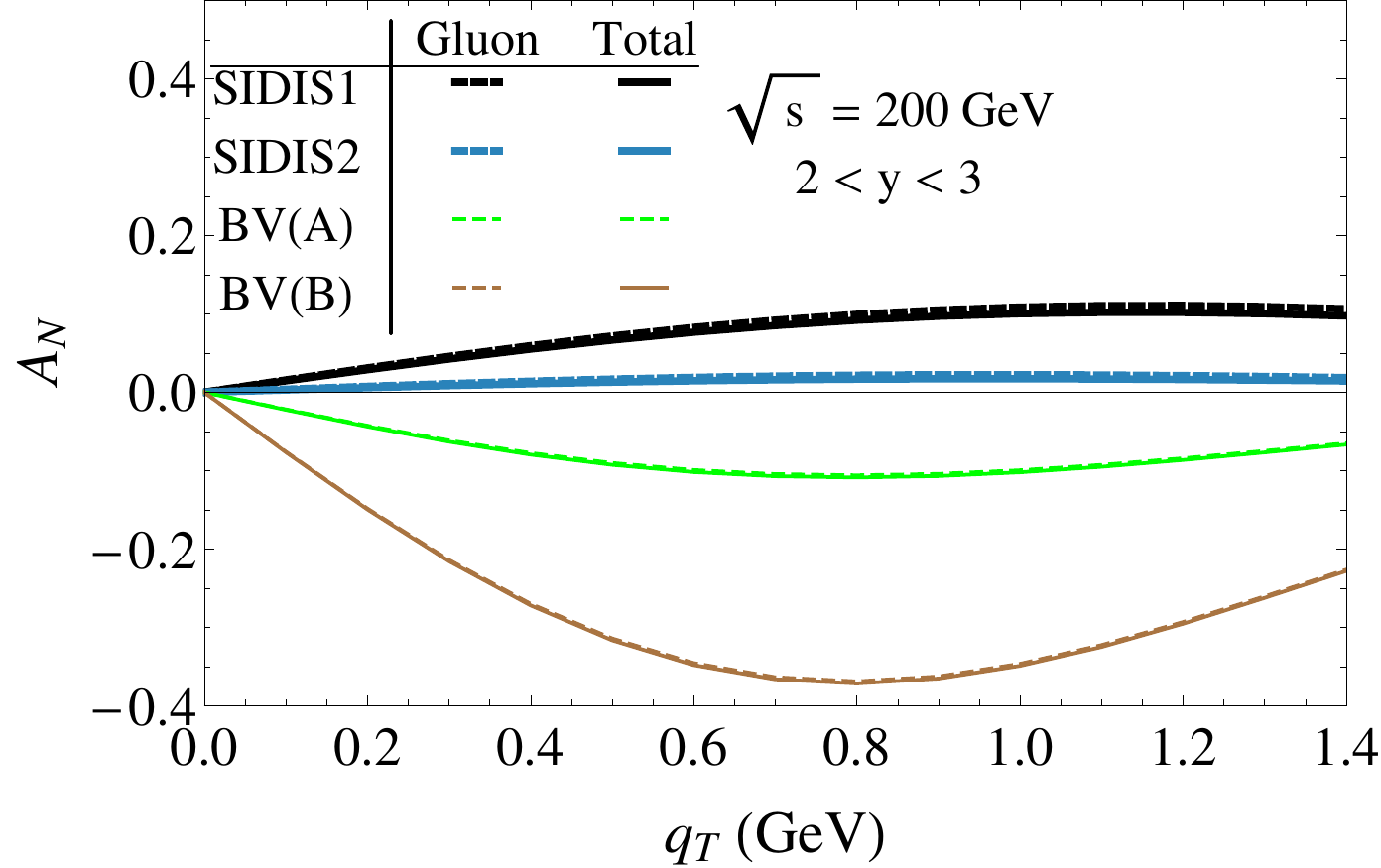}
\includegraphics[width=0.49\linewidth,angle=0]{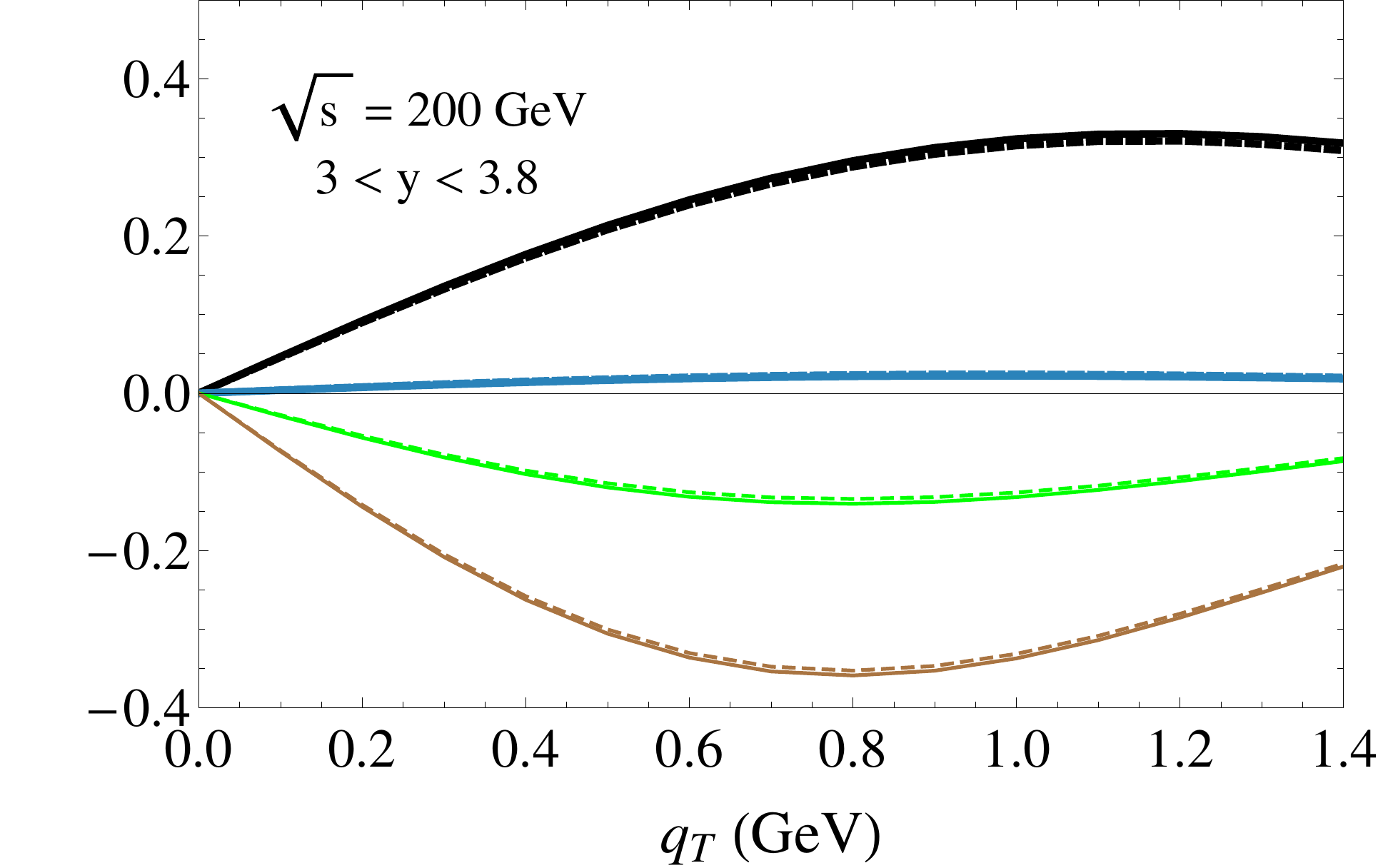}
\caption {Predictions of $q_T $-asymmetry in $p + p^\uparrow \rightarrow J/\psi +X$ at RHIC1 ($\sqrt{s}=$ 200 GeV) energy using DGLAP evolved densities with BV(A), BV(B), DMP-SIDIS1 and DMP-SIDIS2 parameters. Left panel and right panel show $q_T$-asymmetry integrated over the range $2\leq y\leq3$ and $3\leq y\leq3.8$ respectively. The plot shows comparison of total asymmetry (including contribution of both the quark and gluon Sivers functions) with that obtained using only the gluon one. These have been labelled `Total' and `Gluon' respectively.}
\label{p0}
\end{center}
\end{figure}

\begin{figure}[ht]
\begin{center}
\includegraphics[width=0.49\linewidth,angle=0]{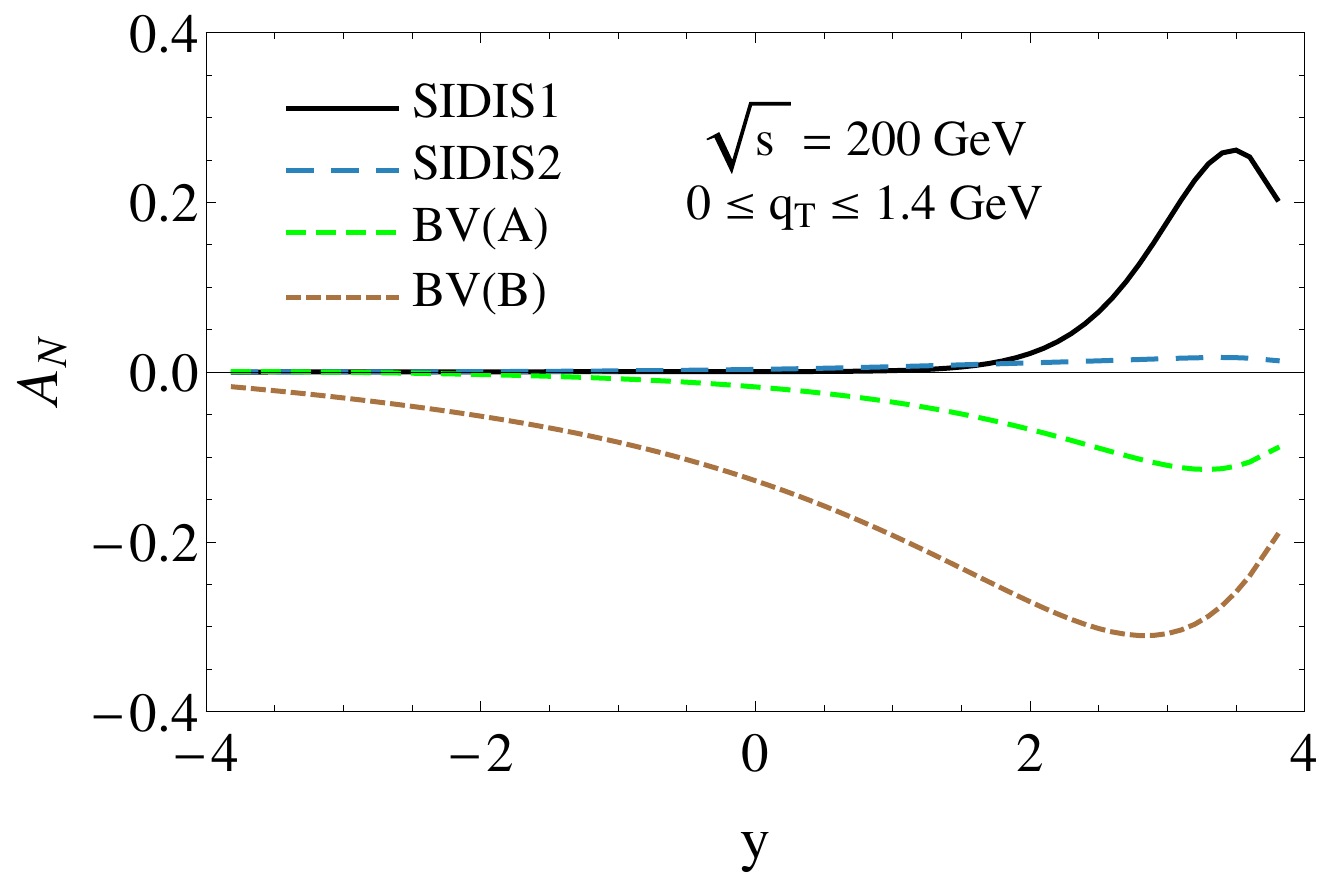}
\caption{Predictions for  $y$-asymmetry in $p + p^\uparrow \rightarrow J/\psi +X$ at RHIC1 ($\sqrt{s}=$ 200 GeV) energy  using DGLAP evolved densities with BV(A), BV(B), DMP-SIDIS1 and DMP-SIDIS2 parameters. $q_T$ integration range is  $0\leq q_T\leq1.4$ GeV.}
\label{p2}
\end{center}
\end{figure}

To assess  the contribution of the QSF over the GSF to the asymmetry, we have compared  total asymmetry (contribution of both quarks and gluon Sivers functions) with the contribution of gluon Sivers function to the asymmetry and found the contribution of  QSF to be negligible in all cases. In Fig.~\ref{p0}, we show this comparison for DMP fits and the BV models of the GSF, at $\sqrt{s}=200$ GeV. We observe that the contribution of  the QSF to the asymmetry  is  indeed very small as compared to contribution of GSF. This assures that the use of this process as a probe of the gluon Sivers function will not be compromised.
In the remaining figures, we have considered contribution from both QSF and GSF.
In Fig.~\ref{p2}, we show rapidity dependence of asymmetry predictions obtained using different sets of DGLAP evolved densities, i.e., the DMP fits and the BV models of the GSF, at $\sqrt{s}=200$ GeV. We observe that the signs of asymmetry obtained using BV parameters and more recent directly fitted DMP parameters are opposite. This is expected as in BV models, the gluon Sivers function is modelled after quark Sivers function and the $d$-quark Sivers parameters have a negative sign as shown in  Tables II and III. Further, the magnitude of asymmetry obtained using the DMP fits is smaller than that obtained using the BV models. Of the two DMP fits,  SIDIS1 gives the larger asymmetry estimates with peak values of about $35\%$ for the $q_T$-asymmetry (with the rapidity range $3\leq y\leq3.8$), and around $26\%$ for the $y$-asymmetry. SIDIS2 on the other hand gives much smaller asymmetries with  peak values of about $2\%$ for the $q_T$-asymmetry (with the rapidity range $3\leq y\leq3.8$) and $2\%$ for the $y$-asymmetry. This large difference in the peak magnitudes of the asymmetry between SIDIS1 and SIDIS2 fits can be understood by looking at the $x_a$-region which contributes to the peaks. For both SIDIS1 and SIDIS2, the peak occurs for $y>3$ which corresponds to the large-$x$ region $x_a\gtrsim0.3$ where the two fits differ greatly in magnitude, as can be inferred from the numbers in Table~\ref{SIDIS-gluon-fits}. It must be mentioned however that the DMP fits do not constrain the GSF very well in this region.

\begin{figure}[ht]
\begin{center}
\includegraphics[width=0.49\linewidth,angle=0]{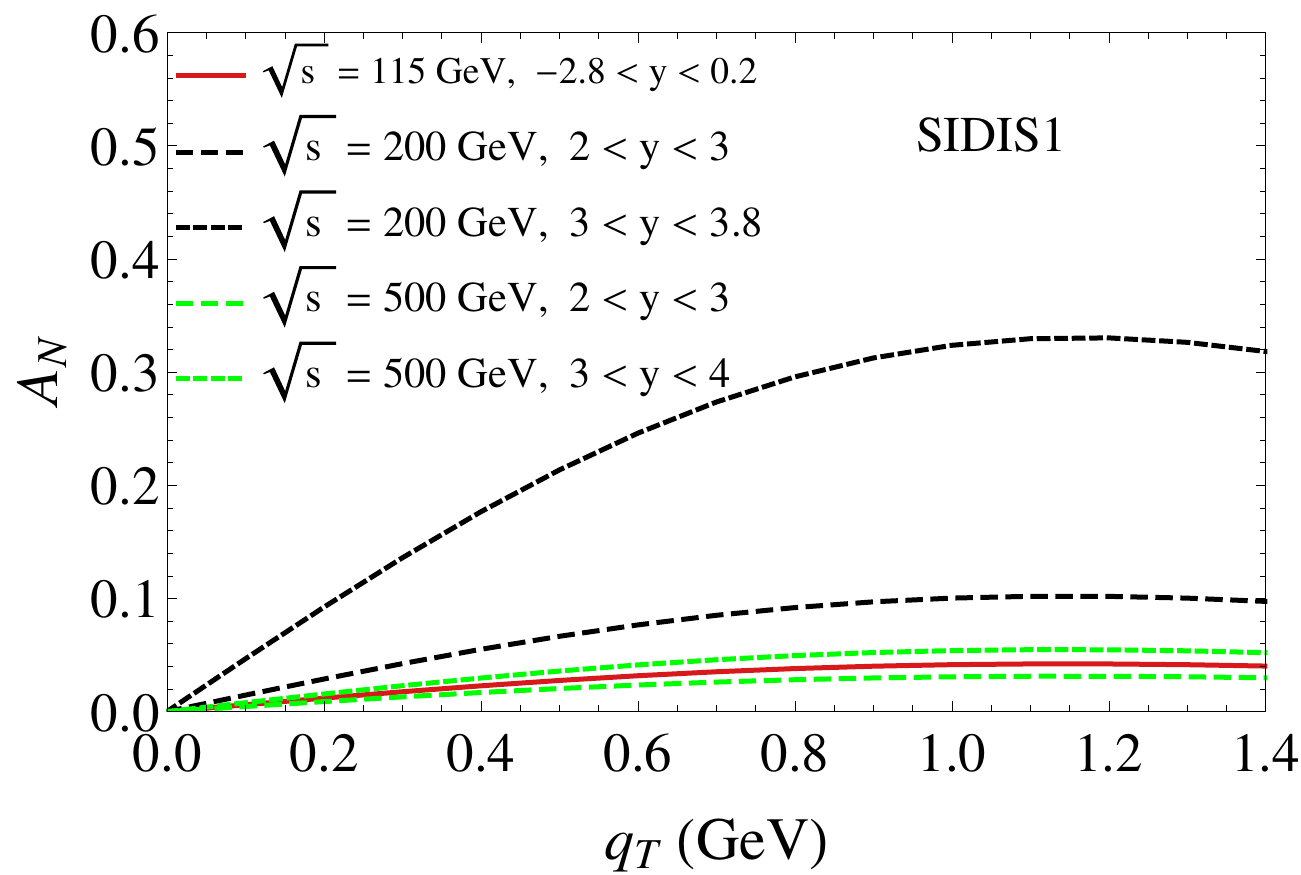}
\includegraphics[width=0.48\linewidth,angle=0]{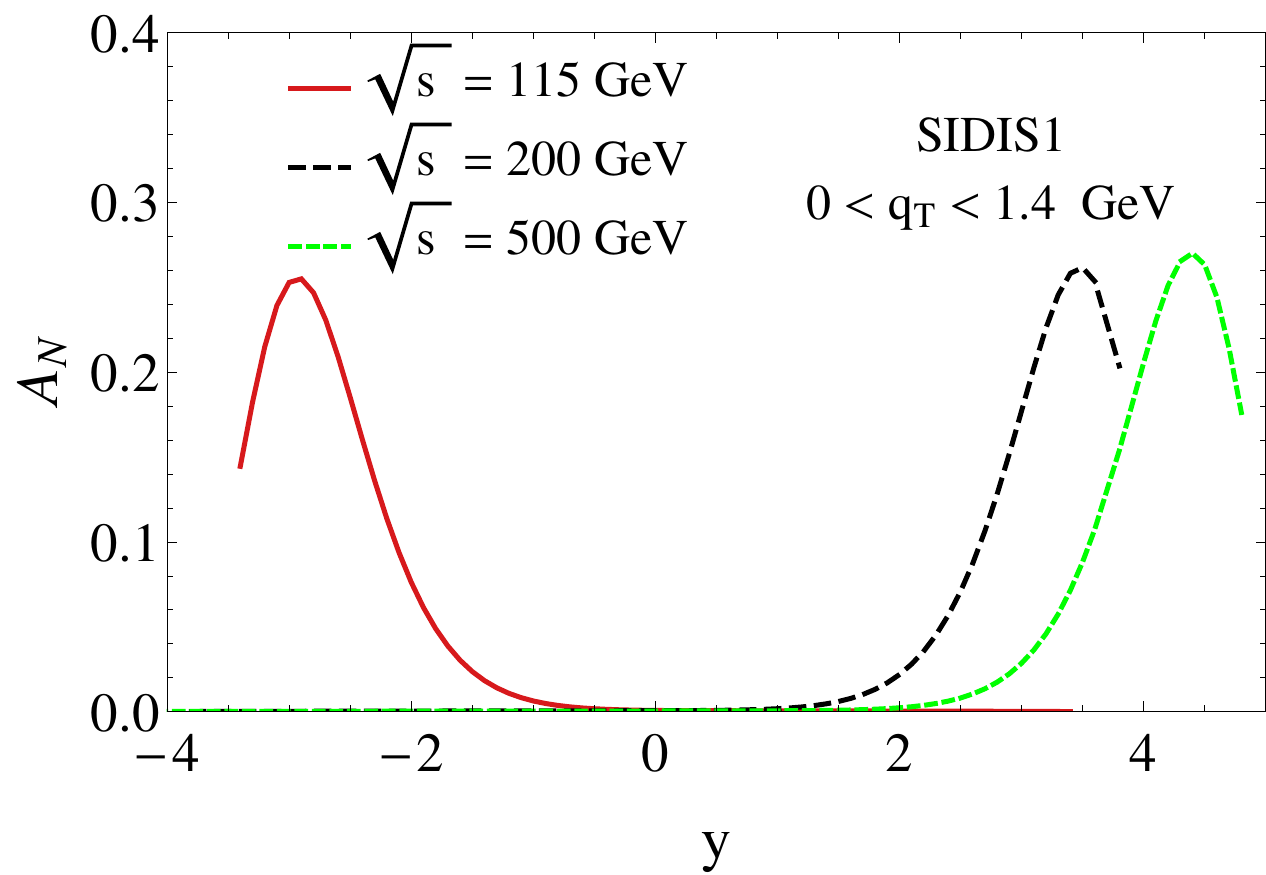}
\caption{Predictions for asymmetry as a function of  $q_T $ (left panel) and y (right panel) obtained using the DMP-SIDIS1 GSF~\cite{D'Alesio:2015uta} parameters for all the three centre of mass values considered ($\sqrt{s}=115$ GeV, 200 GeV, 500 GeV). Integration ranges for rapidity in left panel are $-2.8\leq y \leq 0.2$ for $\sqrt{s}=115$ GeV, $2\leq y \leq 3$  and $3\leq y \leq 3.8 $ for $\sqrt{s}=200$ GeV and $2 \leq y \leq 3 $ and $3 \leq y \leq 4$ for $\sqrt{s}=500$ GeV. In right panel, asymmetry predictions obtained using the DMP-SIDIS1 GSF~\cite{D'Alesio:2015uta} as a function of  $y$ are presented for all three centre of mass values ($\sqrt{s}=115$ GeV, 200 GeV, 500 GeV). Integrati
on range is  for  $q_T $ is $ 0 \leq q_T \leq 1.4 $ GeV in all cases. (Asymmetry peaks in negative $y$ region for AFTER@LHC energy as we have used the convention for fixed target experiments as explained in the Section III)} 
\label{p3}
\end{center}
\end{figure}

\begin{figure}[ht]
\begin{center}
\includegraphics[width=0.5\linewidth,angle=0]{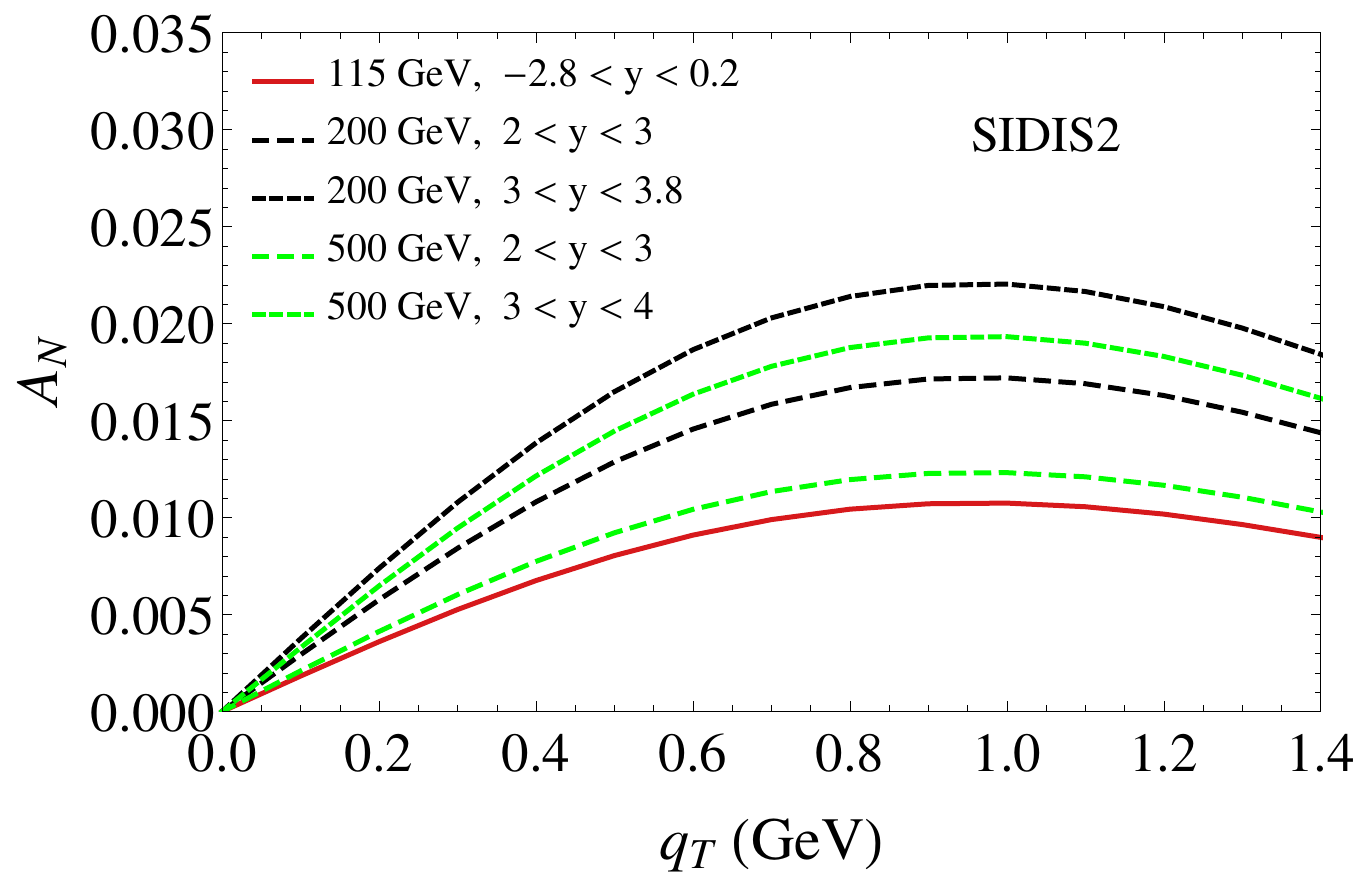}
\includegraphics[width=0.492\linewidth,angle=0]{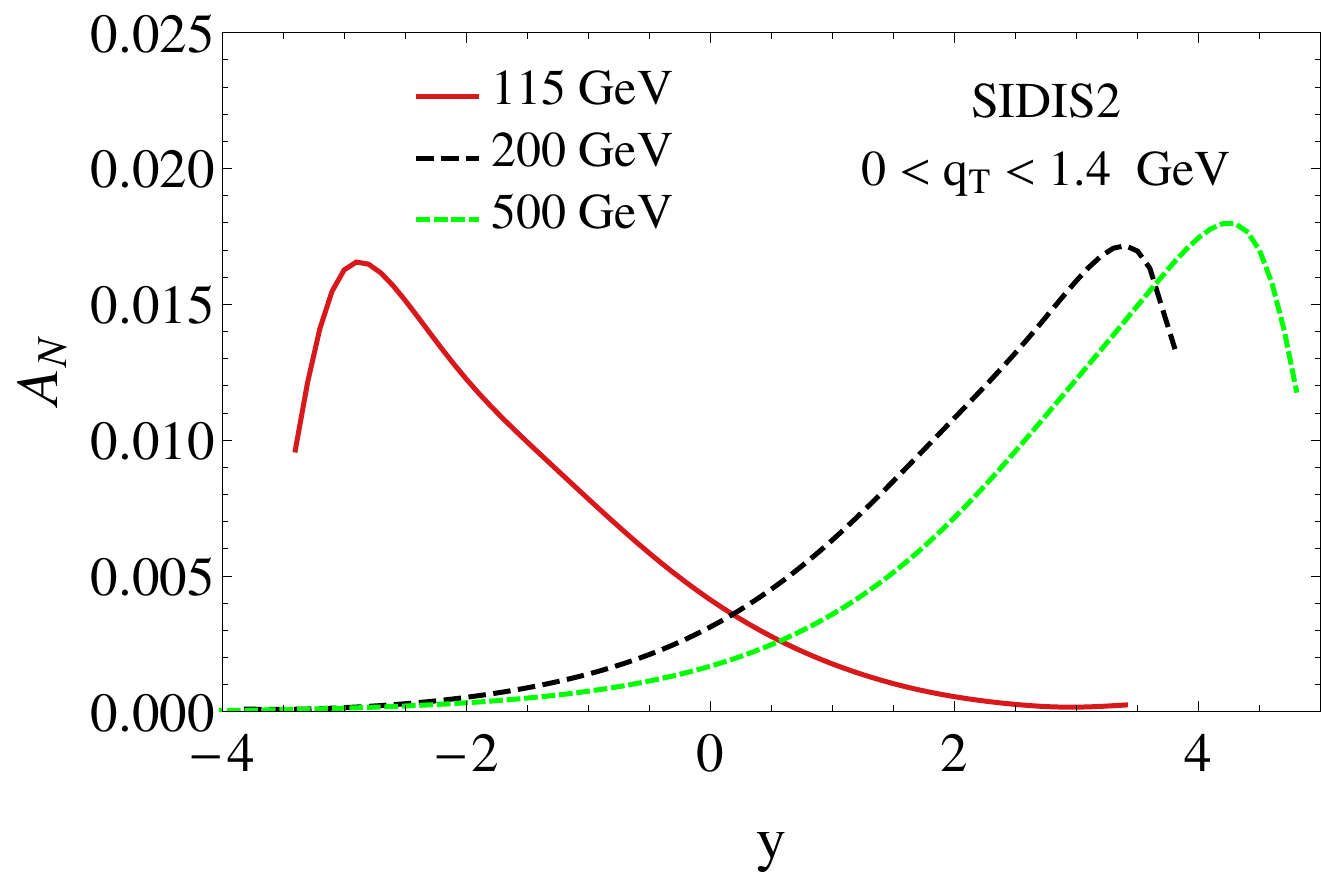}
\caption {Predictions for asymmetry as a function of $q_T$ (left panel) and y (right panel) obtained using the DMP-SIDIS2 GSF~\cite{D'Alesio:2015uta} parameters   for all three centre of mass values considered ($\sqrt{s}=115$ GeV, 200 GeV, 500 GeV). Integration ranges for rapidity in left panel are $-2.8\leq y \leq 0.2$ for $\sqrt{s}=115$ GeV, $2\leq y \leq 3$  and $3\leq y \leq 3.8 $ for $\sqrt{s}=200$ GeV and $2 \leq y \leq 3 $ and $3 \leq y \leq 4$ for $\sqrt{s}=500$ GeV. In right panel, asymmetry predictions obtained using the DMP-SIDIS2 GSF~\cite{D'Alesio:2015uta} are plotted as a function of  $y$ for all three centre of mass values considered ($\sqrt{s}=115$ GeV, 200 GeV, 500 GeV). Integration range for $q_T$ is $ 0 \leq q_T \leq 1.4 $ GeV in all cases. (Asymmetry peaks in negative $y$ region for AFTER@LHC energy as we have used the convention for fixed target experiments as explained in the Section III).}
\label{p4}
\end{center}
\end{figure}

In Figures \ref{p3} and \ref{p4}, we present asymmetry predictions obtained with the DMP fits, SIDIS1 and SIDIS2, for all the three centre of mass energies considered. In both cases, we  find that the $y$-asymmetry scales with $P_L/\sqrt{s}$. It should be noted that in Fig.~\ref{p3}b and \ref{p4}b, the $y$-asymmetry peaks in negative $y$ region for AFTER@LHC c.m energy. This is due to the fact that AFTER@LHC is a fixed target experiment and we have taken $y_{cms}$ to be positive in the (unpolarized) beam direction.  This is in contrast to RHIC1 and RHIC2 curves, where we have used  the convention followed by PHENIX experiment  where rapidity is considered to be positive in the forward hemisphere of the polarized proton. The scaling of $y$-asymmetry with $P_L/\sqrt{s}$ is because the collinear PDFs mostly cancel out between the numerator and the denominator and the $y$-dependence of the asymmetry is mostly determined by the remaining factor $\mathcal{N}_g(x)$, with $x_a$ having a direct correspondence with $y$ (c.f. Eq.~\ref{xy}). This cancellation is of course not absolute, as the integration over the invariant mass $M_{c\bar{c}}$ dilutes the correspondence of $x_a$ with the rapidity and allows the collinear PDFs to affect the $y$-dependence, but we have verified that this effect is small. It must also be mentioned that the assumption that the $k_\perp$ dependence of the TMD is factorized, helps this cancellation as the integrals over the transverse momenta of the partons do not depend on $\sqrt{s}$ and simply produce an overall constant independent of both $y$ and $\sqrt{s}$. We find peak asymmetry values of about $27\%$ with SIDIS1 and $1.9\%$ with SIDIS2.

In the case of the $q_T$-asymmetries, shown in Fig.~\ref{p3}a and \ref{p4}a, we find that the functional form of the $q_T$ dependence remains the same up to an overall factor that depends on $\sqrt{s}$ and the rapidity range. This is also a reflection of the factorized $k_T$-dependence that we have assumed for the TMDPDFs. With SIDIS1, we find that peak asymmetry occurs at $q_T\sim1.1\text{--}1.2$ GeV with peak asymmetry values of 5\%, 33\% and 6\% for $\sqrt{s}=115$ (with $-2.8 \leq y\leq 0.2$) , 200 (with $3\leq y\leq3.8$) and 500 (with $3\leq y\leq4$) GeV respectively, while for SIDIS2 we get substantially lower asymmetries with peak values of 1.1\%, 2.2\% and 1.9\% for $\sqrt{s}=115$ (with $-2.8 \leq y\leq 0.2$), 200 (with $3\leq y\leq3.8$) and 500 (with $3\leq y\leq4$) GeV respectively at $q_T\sim0.9\text{--}1.0$ GeV.

\begin{figure}[ht]
\begin{center}
\includegraphics[width=0.49\linewidth,angle=0]{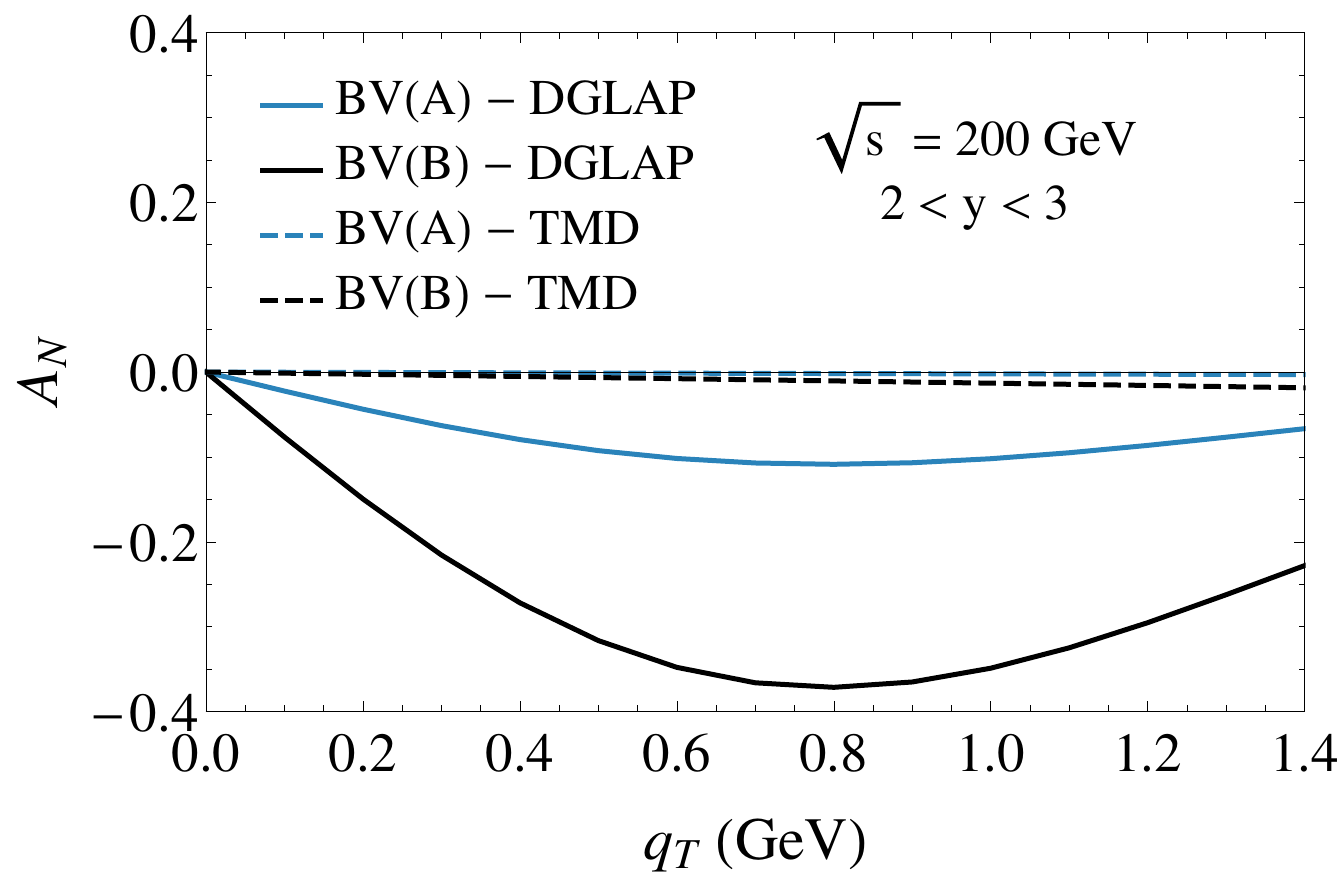}
\includegraphics[width=0.49\linewidth,angle=0]{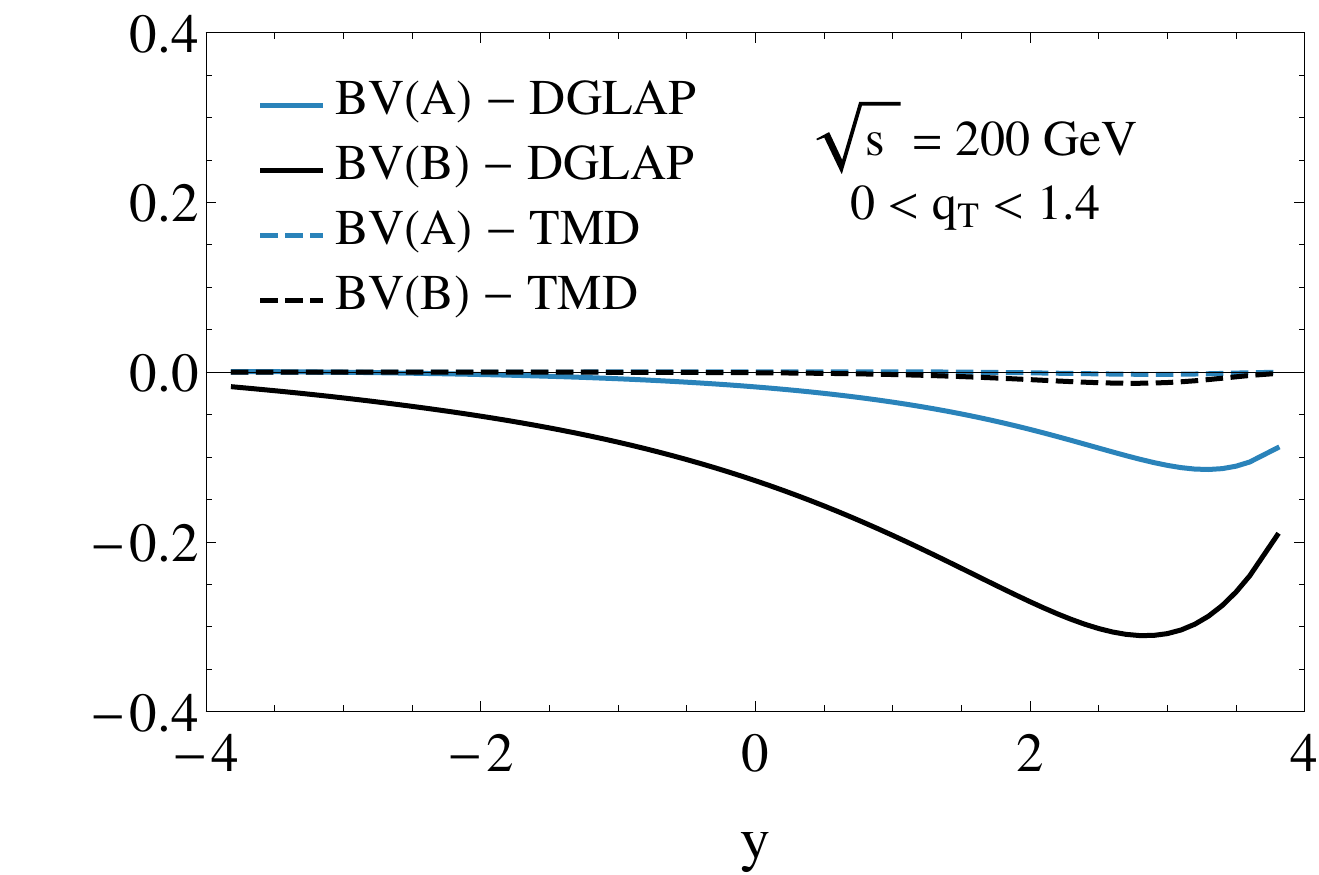}
\caption {Comparison of asymmetry predictions obtained using TMD evolved BV models~\cite{Boer:2003tx} with those obtained using DGLAP evolved BV models. Left panel shows the $q_T$-asymmetry integrated over $2\leq y\leq 3$ and right panel shows the $y$-asymmetry integrated over $0\leq q_T\leq1.4$ GeV.}
\label{p5}
\end{center}
\end{figure}

\begin{figure}[ht]
\begin{center}
\includegraphics[width=0.51\linewidth,angle=0]{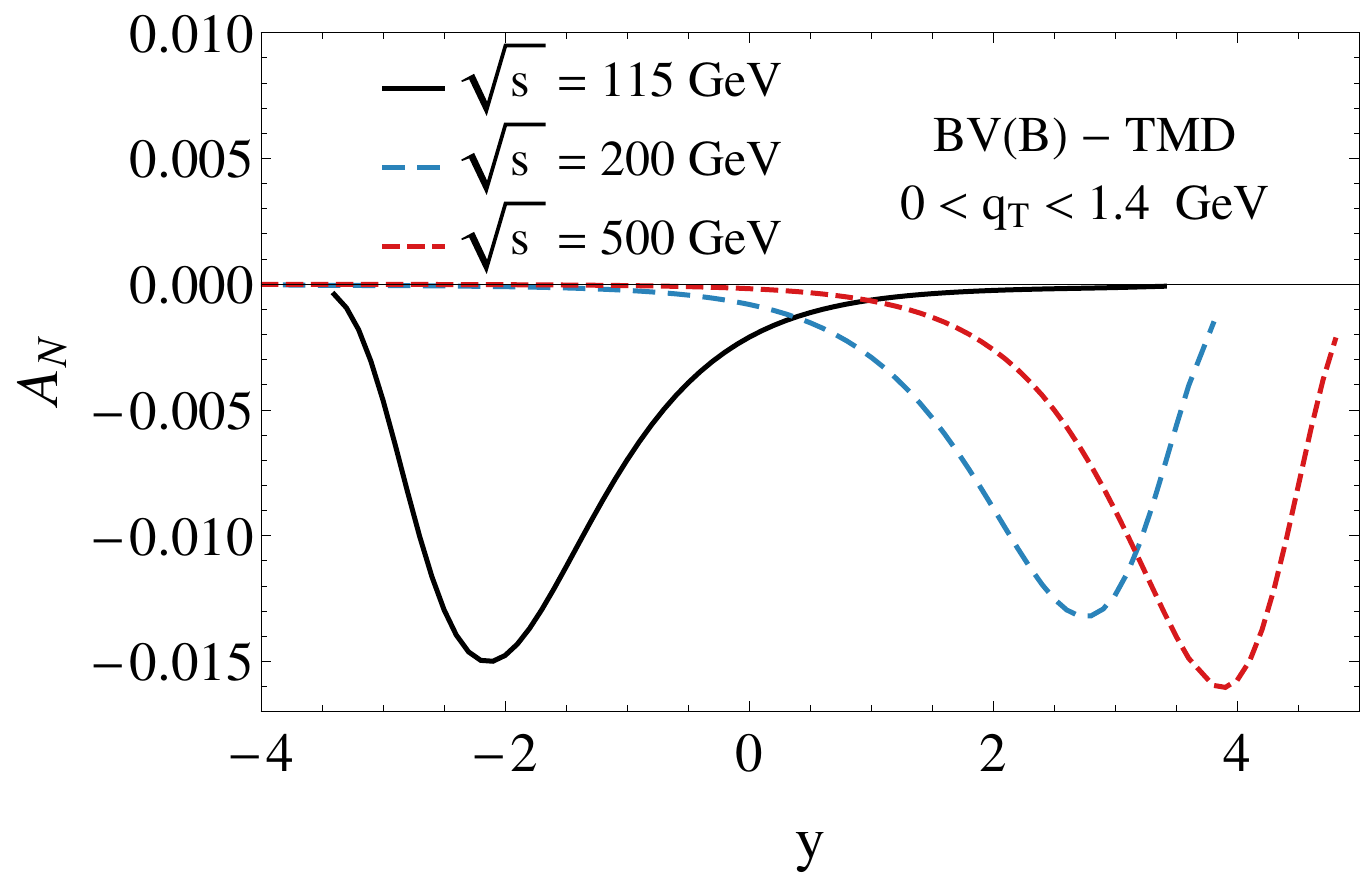}
\caption{Predictions for $y$-asymmetry obtained using the TMD evolved BV (B) model of GSF~\cite{Boer:2003tx}  for all three centre of mass values considered ($\sqrt{s}=115$ GeV, 200 GeV, 500 GeV) with $q_T $  integration range $0\leq q_T\leq1.4$ GeV.}
\label{p6}
\end{center}
\end{figure}

In Figs.~\ref{p5} and \ref{p6}, we look at the effect of TMD evolution on the asymmetry predictions by comparing results obtained with DGLAP evolved densities with those obtained with TMD evolved densities. We do this using the BV (A) and BV (B) models of the GSF (c.f. Section II C). We find that the inclusion of TMD evolution causes the asymmetry predictions to substantially decrease for both models. Furthermore the $P_L/\sqrt{s}$ scaling of the $y$-asymmetry is also affected by TMD evolution as can be seen in Fig.~\ref{p6}, where we show the asymmetries obtained with the TMD evolved BV (B) model, for all three c.o.m energies. While the peak of the asymmetry does shift to larger rapidities with increasing $\sqrt{s}$, the magnitude of the peak varies. This is due to the fact that the $k_T$-dependence of the TMDs is no more factorized, but is instead affected by the $x$-dependence of the collinear PDFs through the $c/b*$ prescription (c.f., Eq.~\ref{fbT},~\ref{fSbT}).

\begin{figure}[ht]
\begin{center}
\includegraphics[width=0.5\linewidth,angle=0]{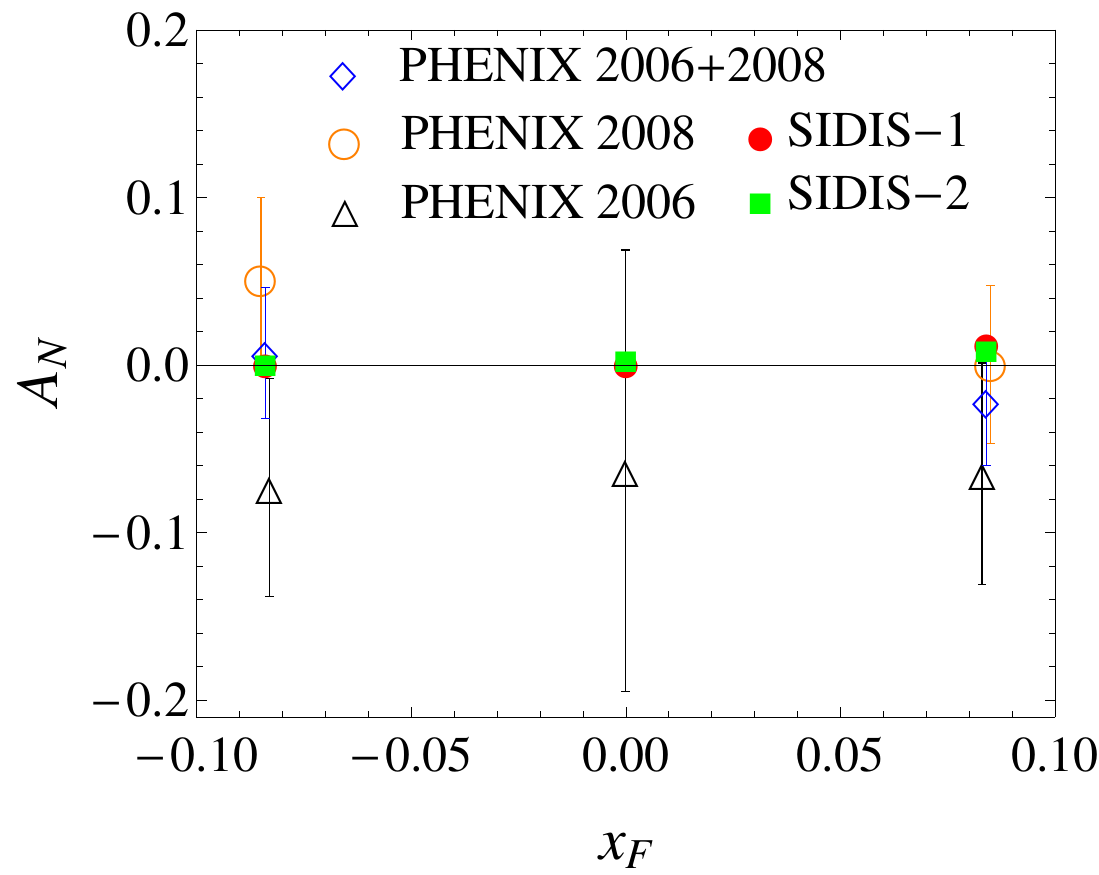}
\caption{Comparison of PHENIX measurements~\cite{Adare:2010bd} of TSSA in  $p + p^\uparrow \rightarrow J/\psi +X$ with predictions obtained using the DMP fits, SIDIS1 and SIDIS2~\cite{D'Alesio:2015uta}. The points for the combined (2006 +2008) data have been offset by 0.01 in $x_F$ for visibility. Asymmetry measurements are in the forward (1.2 $<y<$ 2.2), backward (-2.2$<y<$-1.2) and midrapidity ($|y|<0.35$) regions with $0\leq q_T\leq1.4$ GeV.}
\label{p7}
\end{center}
\end{figure}

\begin{figure}[ht]
\begin{center}
\includegraphics[width=0.49\linewidth,angle=0]{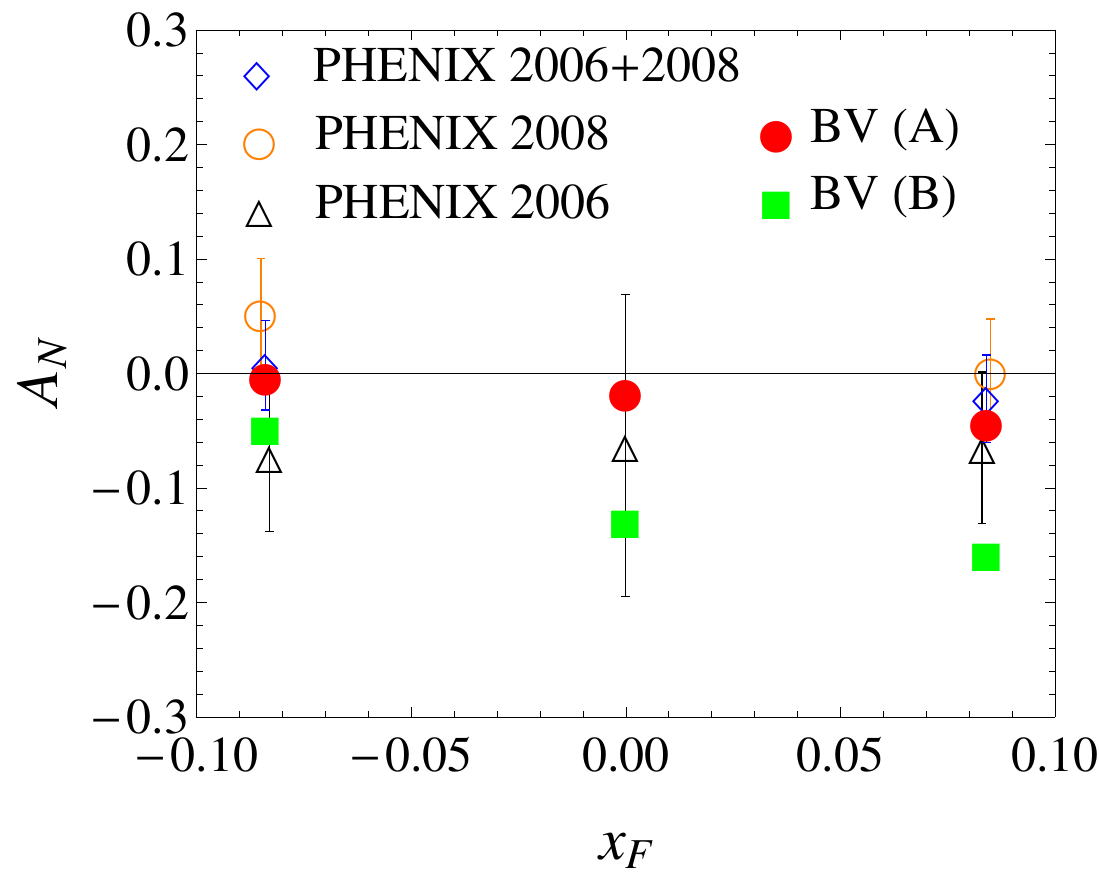}
\includegraphics[width=0.49\linewidth,angle=0]{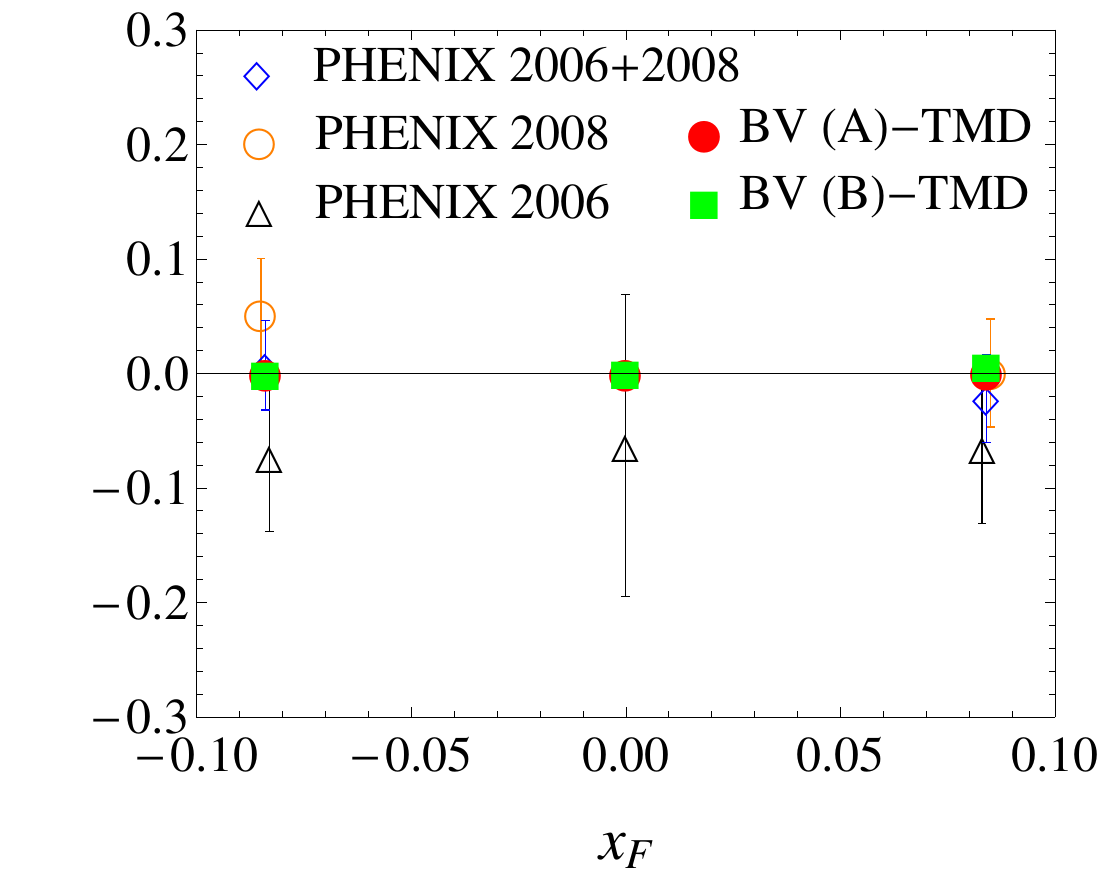}
\caption{Comparison of PHENIX measurements~\cite{Adare:2010bd} of TSSA in $p + p^\uparrow \rightarrow J/\psi +X$ with predictions obtained using BV models of the GSF~\cite{Boer:2003tx}. The points for the combined (2006 +2008) data have been offset by 0.01 in $x_F$ for visibility. Left panel shows comparison with DGLAP evolved BV parameters  and right panel shows the same for TMD evolved BV parameters.}
\label{p8}
\end{center}
\end{figure}

In Figs.~\ref{p7}  and \ref{p8},  we compare our predictions with the asymmetry 
measured at the PHENIX experiment~\cite{Adare:2010bd}. In Fig.~\ref{p7}, we compare the asymmetry predictions obtained using the DMP fits with the data and find that they lie well within the uncertainties. In the forward region, SIDIS1 and SIDIS2 give an asymmetries  of about $1.2 \%$ and $0.9 \%$ respectively. In Fig.~\ref{p8}, we do the same using the DGLAP and TMD evolved BV models. With the DGLAP evolved models, BV (A) gives asymmetries which lie within the uncertainties, whereas BV (B) gives an asymmetry well outside the uncertainties for the forward region. However, the asymmetry predictions obtained with the TMD evolved BV models are substantially smaller than those given by DGLAP evolved BV models and are negligible in all rapidity regions.

\begin{figure}[ht]
\begin{center}
\includegraphics[width=0.6\linewidth,angle=0]{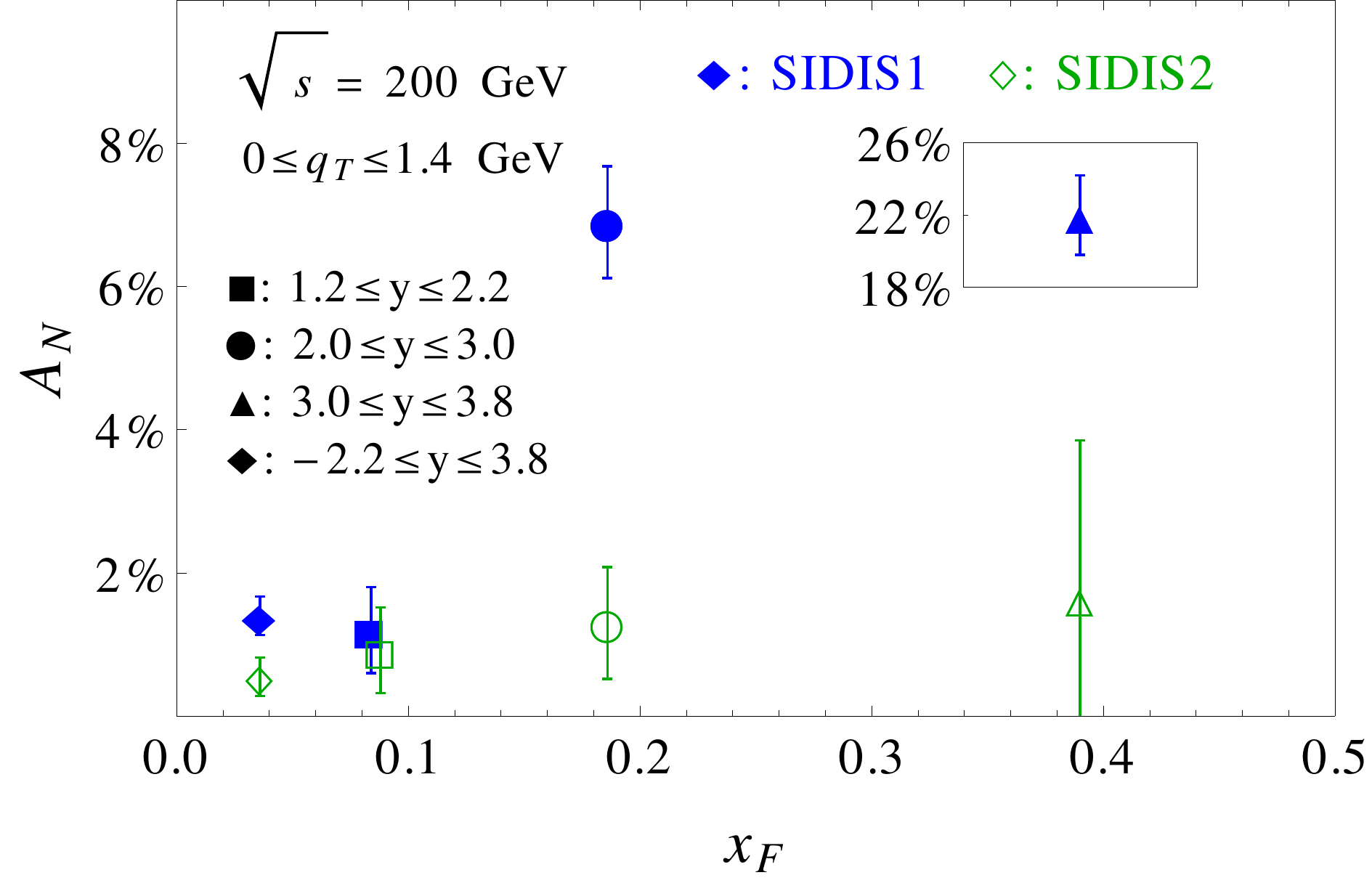}
\caption{Predictions for asymmetry in forward region that would be accessible with the fsPHENIX upgrade~\cite{Barish:2012ha, Aschenauer:2015eha}. Error bars indicate expected statistical errors calculated assuming 1 pb$^{-1}$ of data.}
\label{p9}
\end{center}
\end{figure}

Thus, although the PHENIX results for the asymmetry are compatible with zero, the errors are still large and allow for percentage level asymmetries as given by the DMP fits. Furthermore, they cover a very limited kinematic range and do not rule out larger asymmetries in more forward regions. The proposed fsPHENIX upgrade~\cite{Barish:2012ha, Aschenauer:2015eha} will expand the forward coverage of the detecter to the region $\eta\leq4$. With this in mind, in Fig.~\ref{p9}, we show asymmetry predictions obtained from the DMP fits for the expanded rapidity region that will be covered by the upgrade. The expected statistical error for each point, which is given by $\delta A_N=\sqrt{(1-A_N^2)/N}$, was calculated assuming 1~pb$^{-1}$ of data. Here $N=\sigma^{pp\rightarrow~J/\psi}\times\text{B.R}(J/\psi\rightarrow \mu^+\mu^-)\times F^{\mu^+\mu^-}_{J/\psi}\times\mathcal{L}$, indicates the number $J/\psi$'s that are detected. $\mathcal{L}$ is the integrated luminosity, which we choose to be 1~pb$^{-1}$ here and $F^{\mu^+\mu^-}_{J/\psi}$ is the geometric factor accounting for the planned detector acceptance of leptons: $-2.2\leq\eta_l\leq4.0$.  The cross-section $\sigma^{pp\rightarrow~J/\psi}$ was calculated using the colour evaporation model (CEM), normalised to the total cross-section given in ~\cite{Adare:2006kf}.

In Fig.~\ref{p9}, we would like to highlight the widely differing behavior of the SIDIS1 and SIDIS2 asymmetries with respect to the choice of the rapidity cuts. Note that use of these fits for GSF to make predictions  for our process assumes  TMD factorisation and universality of GSF. These have  yet to be established . Even then these two can be used for demonstration purposes,  as examples of two possible GSF, which have widely different $x$-dependencies. Taking these as examples we demonstrate how a study of asymmetry with different choices of rapidity cuts in the forward region, can probe the $x$-dependence of the GSF. When the whole accessible region $-2.2\leq y\leq 3.8$ is considered, both fits give asymmetries below 2\%, which are similar to each other within $2\sigma$ of the statistical error. For the region $1.2\leq y\leq 2.2$, both fits give asymmetries which are almost indistinguishable given the errors. However for the more forward regions, the two fits give vastly different predictions, with the SIDIS1 fit being much more sensitive to the rapidity cuts. For the region $2\leq y\leq3$, SIDIS1 gives an asymmetry of about 7\%, which is almost five times of the prediction given by SIDIS2. For the region $3\leq y\leq 3.8$, the difference is even larger with SIDIS1 giving as asymmetry of 22\%, which greater than that of SIDIS2 by a factor of 13. This difference is due to the different $x$-dependence of the fits. The region $y>2$ probes the region $x^\uparrow\gtrsim0.11$ where SIDIS1 is much larger than SIDIS2. This difference however, is not seen in the asymmetry without rapidity cuts, i.e., considering the whole region $-2.2\leq y\leq 3.8$, since the cross-section drops rapidly in the large rapidity region. We therefore see that a study of the dependence of the measured asymmetries on the rapidity region over which measurement is made, can give  insight into the $x$-dependence of  the gluon Sivers function. The numerical values of the cross-section, the asymmetries and their associated error are given in Table~\ref{sensitivity}.

\begin{table}[ht]
\centering
\begin{tabular}{|l|l|l|l|l|l|l|}
\hline
~Region~&~$\langle x_F\rangle$ & ~$x^\uparrow$ region~ & ~$\sigma_{J/\psi}\times\text{B.R}(\mu^+\mu^-)$ (nb)~ & ~$\delta A_N$~ & ~$A_N^\text{SIDIS1}$~& ~$A_N^\text{SIDIS2}$~ \\ \hline
~$-2.2<y<3.8~$~& ~0.036~ & ~0.002~-~0.7~ & ~144.0 & ~0.0026~ & ~0.014~ & ~0.0055 \\ \hline
~$1.2<y<2.2$~& ~0.086~ & ~0.05~-~0.14~ & ~29.6 & ~0.0058~ & ~0.012~ & ~0.0092 \\ \hline
~$2<y<3$~& ~0.186~ & ~0.12~-~0.32~ & ~18.2 & ~0.0074~ & ~0.069~ & ~0.013 \\ \hline
~$3<y<3.8$~& ~0.39~ &~0.32~-~0.7~ & ~2.45 & ~0.020~ & ~0.22~ & ~0.017 \\ \hline
\end{tabular}
\caption{Results for cross-section and asymmetry at RHIC1 energy ($\sqrt{s}=200$ GeV), along with expected statistical error (assuming 1 pb$^{-1}$ of data) and approximate $x$-region probed. All numbers given with cuts on lepton rapidity:~$-2.2<y_l<4.0$. Transverse momentum region: $0\leq q_T\leq 1.4$ GeV.}
\label{sensitivity}
\end{table}

Another experiment that might help study the GSF in $pp^\uparrow\rightarrow J/\psi+X$ over kinematic regions not covered by the PHENIX study would be the proposed fixed target experiment AFTER@LHC. This would have a high enough luminosity to make precise measurements of the asymmetry~\cite{Lansberg:2016urh, Anselmino:2015eoa, Brodsky:2012vg, Kikola:2017hnp, Lansberg:2012kf, Rakotozafindrabe:2013au}. Such a fixed target experiment would have a centre of mass energy $\sqrt{s}=115$ GeV with an integrated luminosity of up to 20 fb$^{-1}$ with one year of data taking. In such an experiment the $pp$ centre of mass would be moving with respect to the lab frame with a rapidity $y_\text{cms}=4.8$, allowing large $x$ regions of the target to be probed with the coverage of the ALICE or LHCb detectors. A polarized target would, therefore, offer the possibility of probing the large $x^\uparrow$ ($\gtrsim0.3$) region where the DMP fits do not constrain the GSF. $J/\psi$ production rates obtained with the leading order (LO) CEM indicate that, with an integrated luminosity of 20 fb$^{-1}$, it should be possible to measure the asymmetry with permille precision in the low-$q_T$ region with rapidity range $-3.0\leq y\leq-2.0$ ($1.8\leq y_\text{cms}\leq2.8$). This roughly corresponds to the region $0.2\lesssim x^\uparrow\lesssim0.55$.

\section{Conclusions}
In this paper, we have presented predictions of TSSA in $p + p^\uparrow \to J/\psi + X$ at the RHIC centre of mass energy $\sqrt{s} = 200$ GeV, at which TSSA in $J/\psi$ production has been measured by the PHENIX experiment as well as  at $\sqrt{s} = 115$ GeV, corresponding to the proposed polarized scattering experiment AFTER@LHC and $\sqrt{s} = 500$ GeV, corresponding to the higher RHIC energy using two different parameterizations of gluon Sivers function and in different experimentally accessible kinematic regions. Measurement of TSSA in $J/\psi$ production at PHENIX experiment at RHIC has provided
us an opportunity to compare our predictions with experimental results. 

We have obtained predictions of TSSA using both DGLAP as well as TMD evolved densities.
For the predictions with DGLAP evolved densities, we used recent extractions of GSF from TSSA measurements in $p^\uparrow p\rightarrow\pi^0+X$, referred to here as the DMP parameter sets~\cite{D'Alesio:2015uta}. These extractions of GSF were obtained without taking into account TMD evolution. Hence, we have not used these parameters of GSF to assess the effect of TMD evolution. For the comparison of asymmetries calculated using DGLAP evolved and TMD evolved densities, we have used earlier models of the GSF (referred to as BV models in this work),  which express GSF in terms of $u$ and $d$ quark Sivers function and were  used in our previous work. 

Our results show that the asymmetry obtained using BV parameterization is negative whereas asymmetry 
obtained using SIDIS parameterization is positive. This can be understood considering the fact that
$d$ quark Sivers function has negative sign and in BV model, GSF is written in terms of QSF. The $q_T$ distribution of asymmetry we obtained with BV parameterization is large in magnitude as compared to asymmetry obtained using DMP parameterizations. As far as the $y$ distribution of asymmetry is concerned, the peak magnitudes of asymmetry are similar for all c.o.m energies, with the peak shifting towards larger rapidity values with increasing c.o.m energy.  Comparison of our predictions of asymmetry with PHENIX measurement gives interesting results. When DGLAP evolved TMDs are used, our results with DMP-SIDIS1, DMP-SIDIS2 and BV (A) parameterizations are in agreement with the asymmetry data in forward, backward and midrapidity regions whereas the results obtained using BV (B) parameterization are not within experimental uncertainties of data points. However, when effect of TMD evolution is taken into account, results obtained using BV (B) parameterization also fall within experimental uncertainties.  It may be worthwhile to obtain fits of GSF taking into account TMD evolution and compare predictions of asymmetry obtained using those with the predictions presented here.

As mentioned earlier, since the uncertainties in data points are large, more data will be needed to constrain the GSF. The predictions made in this work are based on the first ever directly fitted parameters of gluon Sivers function and assume a  generalized factorization expression within colour evaporation model of charmonium production. A more detailed analysis  investigating the dependence of our results on charmonium production mechanism, which is still an open question, is under study and will be reported in future. Apart from uncertainties arising from underlying assumptions which include TMD factorization for quarkonium production, which has not yet been established, universality of GSF and choice of a particular production mechanism, another issue that is unresolved so far, there may be further limitations due to restricted region of validity of the  parameter sets used.  However, these studies to understand the GSF and resulting TSSA in $pp^\uparrow$ along with studies probing GSF in open flavour production are expected to play a crucial role in constraining the gluon spin dynamics.
\\

\section{ACKNOWLEDGEMENTS}
We would like to thank J.P.~Lansberg for his very useful comments and suggestions. A.M. and B.S. would like to thank DST, India for financial support under the project no.EMR/2014/0000486 and UGC-BSR under F.7-130/2007(BSR). A.M. would also like to thank the Theoretical Physics Department, CERN, Geneva, where part of this work was done, for their kind hospitality. R.M.G. wishes to acknowledge support from the Department of Science and Technology,
India under Grant No. SR/S2/JCB-64/2007 under the J.C. Bose Fellowship scheme. A.K. would like to thank the Department of Physics, University of Mumbai, for their kind hospitality.

 \bibliography{pp2jpsi.bib}

\end{document}